%% file: paper.tex
\documentclass[conference]{IEEEtran}
\usepackage{fancyhdr} 
\IEEEoverridecommandlockouts
\usepackage[bf,small]{caption} 

\usepackage{hyperref}
\usepackage[hyphenbreaks]{breakurl}
\hypersetup{colorlinks,allcolors=black}
\usepackage[usenames,dvipsnames,svgnames,table]{xcolor}
\hypersetup{
    colorlinks,
   linkcolor=magenta,
   filecolor=magenta,
   urlcolor=magenta,
   citecolor=Blue,
}

\usepackage[normalem]{ulem}
\usepackage{float}
\usepackage{url}
\usepackage{graphicx}
\usepackage{listings}
\usepackage{epstopdf}
\usepackage{color}
\usepackage{subfigure}
\usepackage{xspace}
\usepackage{latexsym}
\makeatletter
\newif\if@restonecol
\makeatother

\usepackage[ruled,vlined,linesnumbered]{algorithm2e}
\usepackage{algorithmic}
\usepackage{enumerate}

\usepackage{makecell}

\usepackage{fontawesome}

\usepackage{wrapfig}
\usepackage{setspace}
\usepackage{blindtext}
\usepackage{amsfonts}
\usepackage{multirow}
\usepackage{pifont}
\usepackage{mathrsfs}
\usepackage{mathtools}
\usepackage{relsize}
\usepackage{ulem}
\usepackage{comment}
\usepackage{footnote}
\usepackage{tikz}
\usetikzlibrary{calc}
\makesavenoteenv{tabular}
\makesavenoteenv{table}
\usepackage{enumitem}

\newcommand\redout{\bgroup\markoverwith
{\textcolor{red}{\rule[0.5ex]{2pt}{0.8pt}}}\ULon}

\newif{\ifSubmit}
\newif{\ifFinal}
\newif{\ifDraft}
\Submittrue
\ifSubmit
\newcommand{\alicomment}[1]{}
\newcommand{\yuecomment}[1]{}
\newcommand{\vcomment}[1]{}
\newcommand{\added}[1]{#1}
\else
\newcommand{\alicomment}[1]{\noindent\textcolor{magenta}{\bf Ali: #1}}

\newcommand{\vcomment}[1]{\textcolor{green}{\textbf{Vasily: #1}}}
\newcommand{\added}[1]{{\color{red}#1}}
\fi

\newcommand{\proj}{\textsc{SFS}}
\newcommand{\fib}{{\small\texttt{fib}}}
\newcommand{\mg}{{\small\texttt{md}}}
\newcommand{\sa}{{\small\texttt{sa}}}
\newcommand{\bench}{FaaSBench}

\newcounter{counter}

\usepackage{booktabs}
\usepackage{siunitx}

\usepackage{zref}
\usepackage{cleveref}
\crefformat{section}{\S#2#1#3}

\newif{\ifrebuttal}

\rebuttaltrue
\ifrebuttal

\fi
\usepackage{balance}

\usepackage[numbers]{natbib}

\begin{document}

\title{{\proj}: Smart OS Scheduling for Serverless Functions}

\makeatletter
\newcommand{\linebreakand}{%
  \end{@IEEEauthorhalign}
  \hfill\mbox{}\par
  \mbox{}\hfill\begin{@IEEEauthorhalign}
}
\makeatother

\author{\IEEEauthorblockN{Yuqi Fu}
\IEEEauthorblockA{\textit{University of Virginia} \\
yfu5@virginia.edu}
\and
\IEEEauthorblockN{Li Liu}
\IEEEauthorblockA{\textit{George Mason University} \\
lliu8@gmu.edu}
\and
\IEEEauthorblockN{Haoliang Wang}
\IEEEauthorblockA{\textit{Adobe Research} \\
hawang@adobe.com}
\linebreakand 
\IEEEauthorblockN{Yue Cheng}
\IEEEauthorblockA{\textit{University of Virginia} \\
yuecheng@virginia.edu}
\and
\IEEEauthorblockN{Songqing Chen}
\IEEEauthorblockA{\textit{George Mason University} \\
sqchen@gmu.edu}
}

\maketitle
\thispagestyle{fancy}
\lhead{}
\rhead{}
\chead{}
\lfoot{\footnotesize{
SC22, November 13-18, 2022, Dallas, Texas, USA
\newline 978-1-6654-5444-5/22/\$31.00 \copyright 2022 IEEE}}
\rfoot{}
\cfoot{}
\renewcommand{\headrulewidth}{0pt}
\renewcommand{\footrulewidth}{0pt}

\input{abstract}
\input{introduction}

\input{background}

\input{rte}

\input{motivation}

\input{design}
\input{implementation}

\input{evaluation}

\input{discussion}

\input{related}

\input{conclusion}

\vspace{-4pt}
\section*{Acknowledgments}
\vspace{-2pt}

We are grateful to the anonymous reviewers for their valuable comments and suggestions that improved the paper. This work is sponsored in part under an NSF CAREER Award CNS-2045680, CCF-1919075, CCF-1919113, OAC-2106446, CMMI-2134689, CNS-2007153, an Adobe Research gift, and an AWS CloudBank grant.

\label{startofrefs}
\clearpage
\newpage
\balance



{
\bibliographystyle{plain}
\bibliography{refs}
}
\clearpage
\newpage

\end{document}
\endinput

%% file: abstract.tex
\begin{abstract}
Serverless computing
enables a new way of building and scaling cloud applications by allowing developers to write fine-grained serverless or cloud functions. The execution duration of a cloud function is typically short---ranging from a few milliseconds to hundreds of seconds. 
However, due to resource contentions caused by public clouds' deep consolidation, the function execution duration may get significantly prolonged
and fail to accurately 
account for 
the function's true resource usage.
We observe that the function duration can be highly unpredictable with huge amplification of 
more than $50\times$ for an open-source FaaS platform (OpenLambda).
Our experiments show that the OS scheduling policy of cloud functions' host server can have a crucial impact on performance. \added{The default Linux scheduler, CFS (Completely Fair Scheduler), being oblivious to workloads, frequently context-switches short functions, causing a turnaround time that is much longer than their service time.}
 
We propose {\proj} (Smart Function Scheduler), \added{which works entirely in the user space and carefully orchestrates existing Linux FIFO and CFS schedulers} to approximate Shortest Remaining Time First (SRTF). 
{\proj} uses two-level scheduling that seamlessly combines a new FILTER policy with
Linux CFS, to trade off increased duration of long functions for significant performance improvement for short functions. We implement {\proj} in the Linux user space and port it to OpenLambda. Evaluation results show that {\proj} significantly improves short functions' duration with a small impact 
on relatively longer functions, compared to CFS. 
\end{abstract}

%% file: introduction.tex
\vspace{-4pt}
\section{Introduction}
\vspace{-2pt}

Serverless computing, or Function-as-a-Service (FaaS), 
enables a new way of building and scaling applications and services by allowing developers to break traditionally monolithic server-based applications into finer-grained cloud functions. Developers write function logic while the service provider performs the notoriously tedious tasks of provisioning, scaling, and managing the backend servers~\cite{gray_stop} that the functions run on. 
Serverless computing solutions are growing in popularity and finding their way into both commercial clouds (e.g., AWS Lambda~\cite{lambda}, Azure Functions~\cite{azure_func}, and Google Cloud Functions~\cite{google_func}, etc.) and open-source projects (e.g., OpenLambda~\cite{openlambda_hotcloud16, openlambda}, OpenWhisk~\cite{openwhisk}). Popular uses of serverless computing today are event-driven and stateless applications such as web/API serving, image processing, and batch ETL (extract, transform, load)~\cite{serverless_usecases}.

The execution duration of a cloud function is typically short--ranging from a few milliseconds (ms) to a few seconds~\cite{serverless_in_the_wild}. Therefore, 
FaaS providers charge users at a fine granularity. For example, AWS Lambda bills on a per invocation basis ($\$0.02$ per 1 million invocations) and charges the usage of bundled CPU-memory resources by rounding up the function's execution duration to the nearest 1 ms with a rate of $\$0.0000166667$ per second for each GB of memory.

This fine-grained pricing model would be advantageous and fair to FaaS users if the execution duration of a function does not vary much (ideally one would expect that to be equivalent to the turnaround time as if the function was executed on a dedicated machine). This is particularly important considering  
the short-lived \added{and highly heterogeneous} nature of cloud functions: 
\added{a majority of cloud functions have short execution duration while the execution times of all functions span seven orders of magnitude (from ms to hundreds of seconds).}
However, due to resource contentions caused by public clouds' deep consolidation, function execution duration---the turnaround time that measures the time when a function 
starts execution till the time when the function finishes execution and returns---gets prolonged and fails to accurately
account for
the actual resource usage of a successfully finished function. 
This covertly leads to overcharges to the users and potentially making them game the system in the long run~\cite{ginzburg2020serverless}.  

Admittedly, function execution duration amplification may be caused by contentions from various levels of resources including CPU cache, CPU, memory, and network. 
However, our study shows that the CPU scheduling policy of host machines, e.g., the widely-used Linux CPU scheduling policy, Completely Fair Scheduler or CFS,
can have a crucial impact on the execution duration of cloud functions hosted therein, therefore, a function scheduler must incorporate the unique FaaS workload patterns. 

How to mitigate this amplification for short-job-dominant FaaS workloads is an open challenge, which, to the best of our knowledge, has not been well investigated.
On a similar note, there are no well-defined performance SLOs (service level objectives) for short-job-dominant FaaS applications; 
one potential example SLO can be: ``{\it $X\%$ of function invocations must be finished within a soft/hard-bounded ratio with respect to the duration that this function would observe if running in an ideally isolated environment}''. 

\emph{Our key observation in this paper is that a majority of cloud functions in production FaaS workloads are short-lived with a wide spectrum of execution duration; the default Linux scheduler, CFS, frequently context-switches short functions, causing unfairly long waiting time, and therefore, longer turnaround time than they should have.} 

\added{CFS is general-purpose and workload-oblivious, attempting to achieve CPU-task-level fairness: CPU tasks, no matter long-running or short-lived, get proportional share of the CPU resource under fine-grained time slices. 
This causes all CPU tasks with same priority to spend ``fair'' amounts of time waiting to be rescheduled. This inevitably leads to severely imbalanced function \emph{run-time effectiveness (RTE)},
a new efficiency metric that we define to capture the ratio of function service time (aggregate CPU time) to end-to-end turnaround time (sum of the aggregate CPU time and the waiting time).}

This motivates us to adopt Shortest Remaining Time First (SRTF)---\added{a preemptive version of shortest job first (SJF)---which always schedules jobs that will complete the quickest}.
However, it is impossible to directly apply SRTF as it is an offline algorithm.
To this end, we present {\proj}, a user-space function scheduler that minimizes turnaround time for short-function-intensive FaaS workloads. 
\added{{\proj} works entirely in the user space, leveraging existing kernel scheduling policies (FIFO and CFS) to approximate SRTF. 
For this purpose,
{\proj} adopts two-level scheduling:}
at the top level, {\proj} uses a new FILTER (FIFO-like) algorithm that schedules functions in the order they are enqueued and preempts them if they do not finish in a dynamically changing time slice; 
at the bottom level, those filtered functions
from the top level continue in Linux CFS.
This way, short functions 
can execute in their entirety without any context switch, or with minimum context switches if needed, in order to finish faster. 
The objective is to minimize the function execution duration and maximize the \added{RTE metric} such that the ``pay-per-use'' promise is delivered and unfair overcharges are reduced.

\added{{\proj} presents a novel and practical user-space scheduling solution that bridges the divide between custom, user-space scheduling and kernel scheduling: existing OS scheduling is FaaS-workload-oblivious and thus affects function performance; {\proj} utilizes historical workload statistics obtained in the user space to make informed scheduling decisions by automatically steering underlying OS scheduling policies.}
{\proj} strikes a balance between \added{waiting time} and request service time.
{\proj} is transparent to existing FaaS platforms and requires minimum modifications for them to use {\proj}.
{\proj} is also OS-scheduler-agnostic and does not require kernel modifications.

In summary, this paper makes the following contributions:
\begin{itemize}[noitemsep,leftmargin=*]
    \item  \added{Through a performance characterization study on an open-source FaaS platform (OpenLambda), we identify efficiency problems of existing Linux schedulers (CFS, FIFO, and RR) on serverless function scheduling.}

    \item We design a new scheduler, {\proj}, which approximates SRTF.
    {\proj} \added{features a novel FILTER algorithm in the user space that dynamically steers existing OS schedulers based on workload patterns to enable more efficient scheduling for FaaS workloads.}

    \item We implement {\proj} as a standalone, user-space scheduler that can be easily ported with existing FaaS platforms. 

    \item We perform extensive evaluation on standalone {\proj} and an {\proj}-ported OpenLambda. Results show that {\proj} significantly outperforms
    CFS: {\proj} improves turnaround time of short functions by two orders of magnitude against CFS with very small user-space overhead. 
    
\end{itemize}    

SFS targets the overall performance of a majority of functions that are short-lived. 
Experimental results show that {\proj} improves the execution duration of $83\%$ of the functions by $49.6\times$ on average compared to CFS; for the remaining $17\%$ of the functions that are relatively longer,
they run $1.29\times$ longer on average under {\proj} than CFS. 
{\proj} is open sourced and publicly available at \url{https://github.com/ds2-lab/SFS}.

%% file: background.tex
\vspace{-6pt}
\section{Background}
\label{sec:background}
\vspace{-2pt}

\subsection{FaaS Overview}
\vspace{-2pt}

Serverless computing handles virtually all system administration tasks, making it easier for users to deploy and scale their cloud applications and services~\cite{berkeley_serverless}. FaaS providers offer a flexible interface for defining cloud functions, which allows developers to focus on core application logic using languages such as Python, JavaScript, Java, Go, and others. FaaS providers in turn auto-scale function executions in a demand-driven manner, hiding tedious server configuration and management tasks from the users. 

Cloud functions are deployed and executed in virtualized environments such as containers or virtual machines (VMs) for isolation and safety. 
A typical workflow of function deployment and execution works as follows.
Step~1: A user submits the function code (via either a web interface or packaged .zip/container image files) to the FaaS platform for function creation.
Step~2: The user executes the created function by sending an HTTP invocation request to a FaaS scheduler.
Step~3: the FaaS scheduler forwards the invocation request to a FaaS worker that is running on a  
resource-rich host machine.
Step~4: The FaaS worker creates a virtualized environment and installs the necessary dependencies for the virtualized environment before the function can be started.
Step~5: After all previous steps are successfully completed, the FaaS worker sends the function request to the host OS, which in turn starts the function execution as an OS process. 

While there are already extensive studies focusing on reducing functions' cold startup penalty (Step~2-4 in last paragraph) in  FaaS~\cite{serverless_in_the_wild, sand_atc18, sock_atc18, agile_cold_start_hotcloud19, catalyzer_asplos20, faascache_asplos20}, in this paper, we aim to fill the missing gap by focusing on the ``\emph{last mile}'' efficiency of function execution, i.e., \emph{OS scheduling}, in Step~5.

\vspace{-6pt}
\subsection{OS Task Scheduling}
\label{subsec:sched_overview}
\vspace{-2pt}

Cloud functions are eventually scheduled and executed by a host OS. 
Functions typically have a short execution duration and small CPU-memory footprint, making FaaS workloads increasingly consolidated. For example, a large bare-metal machine with 96 CPU cores, 384 GB of memory, and multi-TBs of NVMe SSDs can easily host tens of thousands of, if not more, function instances~\cite{firecracker_nsdi20}. This statistical multiplexing makes it feasible for a FaaS provider to execute thousands of function processes concurrently on a single host.

\noindent\textbf{Basic Scheduling Policies.}
First in, First out (FIFO) and Round-Robin (RR) are among the most basic scheduling policies.
They have different tradeoffs. 
When a FIFO task starts running, it runs to completion\footnote{Modern OSes must handle sophisticated situations such as I/O and priority. A FIFO task, once started, continues to run until it voluntarily yields control over CPU, blocks,
or is preempted by a higher priority CPU task.}. 
Similar to FIFO, core-granular scheduling~\cite{granular_socc19} designates a single core to a function and allows it to run to completion. However, also like FIFO, core-granular scheduling may hurt response time when the system is highly consolidated and under high utilization with a line of queued tasks. 
RR can be used to optimize responsiveness.
RR runs a CPU task for a time slice and then switches to the next queued task. However, since the execution of a CPU task is divided into multiple slices, RR sacrifices turnaround time.

\noindent\textbf{Proportional-Share Scheduling.}
Proportional-share scheduling is a type of CPU scheduling algorithms 
commonly used by today's OSes including VM scheduling and Linux scheduling. Proportional-share scheduling focuses on fairness and  attempts to guarantee that each CPU task
obtains a certain percentage of CPU time based on the task's priority. Well-known examples of proportional-share scheduling include lottery scheduling~\cite{lottery_osdi94}, Xen's credit scheduler~\cite{xen_sosp03, xen_credit}, and Linux's default scheduler CFS. 
CFS is the \emph{de facto}, and the most commonly-used open-source OS task scheduler in productive environments including public clouds~\cite{aws, gcp, azure} and companies~\cite{ghost_sosp21, cfs_bandwidth}.

Given its popularity and prevalence, we choose Linux's general-purpose CFS scheduler as a baseline and briefly describe how it works. In fact, the two virtualization techniques commonly used by today's FaaS platforms, containers 
and KVM-based VMs, both rely on CFS for OS task scheduling. For example, Docker containers~\cite{docker} are used by open source FaaS platforms such as OpenLambda~\cite{openlambda_hotcloud16}, OpenWhisk~\cite{openwhisk}, and OpenFaaS~\cite{openfaas}, while AWS Lambda's Firecracker microVM~\cite{firecracker_nsdi20} uses KVM for managing Lambda functions. 

\noindent\textbf{Linux CFS.}
CFS proportionally divides the physical time into fine-grained time slices among all CPU tasks based on their weights (priorities). 
CFS tracks the CPU time usage of each task using a virtual runtime ({\small\texttt{vruntime}}) scheme. 
\added{{\small\texttt{vruntime}} records the CPU time that a CPU task has used weighted by its priority.} 
In a multi-core system, each physical CPU core has its own {\small\texttt{runqueue}}, which is a red-black (RB) tree ordered by {\small\texttt{vruntime}}. A task will first be assigned by CFS to a {\small\texttt{runqueue}}; the task's location in the {\small\texttt{runqueue}} RB tree determines roughly when in the future it can execute; at each scheduling tick (i.e., the end of a time slice), CFS picks the next task that has the smallest {\small\texttt{vruntime}} from the RB tree. As the FaaS workload is increasingly consolidated, it is common to have thousands of concurrently running function processes that multiplex the limited amount of CPU cores on the host machine. Therefore, 
the weight of a task simply indicates a \emph{relative} CPU share, but not an absolute CPU share that a user would expect the function to get based on the function's resource configuration. 
\added{Once preempted, the task needs to \emph{wait} in the {\small\texttt{runqueue}} for its next turn to run. While waiting, the task' {\small\texttt{vruntime}} does not tick.}

%% file: rte.tex
\section{Why is CFS a Poor Match?}
\label{sec:rte}

\noindent\textbf{Run-time Effectiveness (RTE).}
\added{CFS is a poor match for the emerging, short-function-intensive FaaS workloads, which values turnaround time. The fundamental mismatch comes from CFS' lack of workload awareness:
CFS ensures a fair proportion of CPU time to all the CPU tasks but
does not distinguish if a CPU task is long-running or short-lived. 
However, this application-level knowledge is critical to application performance, especially if applications mostly consist of short jobs, e.g., a FaaS workload~\cite{serverless_in_the_wild}. 
Under a proportional-share scheduler such as CFS, the \emph{fairness} is defined as follows: within a given time interval, all CPU tasks, if with the same priority, are assigned the same amount of CPU time to execute. We argue in this work, while it is ``fair'' to all CPU tasks from the low-level OS perspective, such ``fairness'' may inevitably create \emph{unfairness} to the user-level applications -- in our case these contained in the FaaS workload, since the waiting time may be disproportional to the execution time, considering the execution time diversity of FaaS workloads. In fact, 
FaaS workloads have unique characteristics that make existing Linux's ``fair'' scheduler actually \emph{unfair}. Cloud functions feature a long spectrum of execution duration (\cref{subsec:moti_azure}). Short functions with an execution duration of several ms to tens of ms are more sensitive to waiting time
than longer functions that execute for, say tens of seconds. A mixture of such short 
and long functions 
co-located in the same server could spend roughly equal amounts of time waiting in {\small\texttt{runqueues}} before those short functions finish, resulting in disproportionally long waiting time. 
To quantify this affect, 
we define a new efficiency metric in this paper, function Run-Time Effectiveness (RTE) as follows:
\begin{equation}
    RTE = \frac{\sum CPU^i}{turnaround time}
\end{equation}
where $CPU^i$ means the CPU time allocated to this function in the $i^{th}$ round before the function returns.
Thus, effectively, RTE reflects the ratio of the service time to the turnaround time. An RTE of 1 is the theoretically highest that a function could achieve, meaning that the function runs to completion without being preempted, so higher scores are better.} 
Furthermore, RTE also indicates if a FaaS user has been overcharged. The closer an RTE to 1, the less overcharges that a user had to pay. For ideally CPU-intensive functions, an RTE of 1 indicates that the user is not overcharged at all. However, one should note that, in practice, it is not common to have functions with pure CPU bursts; therefore, an RTE score achieved under zero interference, though smaller than 1, would still represent a best-case baseline for comparison purposes.

\noindent\textbf{Tradeoffs.}
\added{Efficiently scheduling short and long jobs  is a decades-old problem~\cite{light_tail_or12,overload_tit06, mlfq_sigmetrics95, bsd_os_txt, size_sched_tocs03, srtf_mor01, shinjuku_nsdi19, shenango_nsdi19}. Long jobs' performance will get affected under priority scheduling that approximates SRTF. The challenge is how to balance the tradeoff between the performance improvement for short jobs and performance loss for long jobs. We revisit this problem from a new angle---minimizing severely \emph{disproportional waiting time} for short functions by trading off \emph{disproportionally-increased turnaround time} for long functions---in the context of serverless function scheduling. That is,
{\proj} aims to trade a smaller impact on long functions for significant performance improvement for short functions, a huge win for the short ones and a much smaller (relative) penalty for the longer ones.}

%% file: motivation.tex
\begin{figure}[t]
\begin{center}
\includegraphics[width=0.35\textwidth]{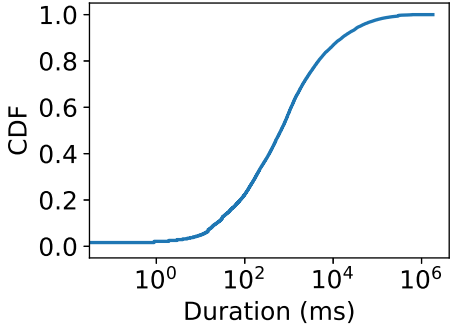}
\vspace{-10pt}
\caption{
CDF of the average function execution duration of Azure Functions traces.
}
\label{fig:moti_azure_cdf}
\vspace{-20pt}
\end{center}
\end{figure}

\begin{figure*}[t]
\begin{center}
\subfigure[Execution duration distribution.] {
\includegraphics[width=.37\textwidth]{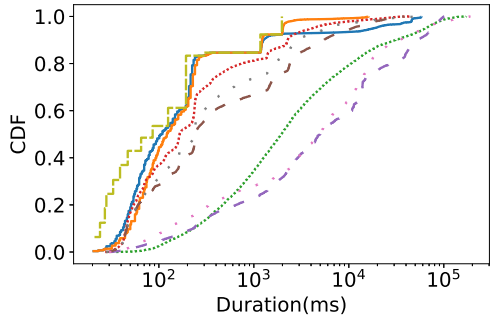}
\label{fig:moti_perf}
}
\hspace{30pt}
\subfigure[Run-time effectiveness (RTE) distribution.] {
\includegraphics[width=.37\textwidth]{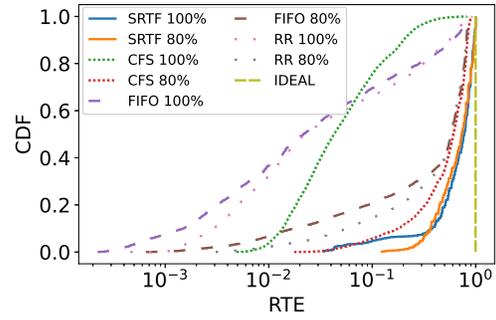}
\label{fig:moti_rte}
}
\vspace{-5pt}
\caption{
Performance and RTE of an Azure-sampled workload on OpenLambda with different scheduling policies and different loads. 
}
\label{fig:moti}
\end{center}
\vspace{-15pt}
\end{figure*}

\vspace{-2pt}
\section{Motivation}
\label{sec:motivation}

\vspace{-2pt}
\subsection{Azure Functions Workloads}
\label{subsec:moti_azure}
\vspace{-2pt}

The Azure Functions workload datasets~\cite{serverless_in_the_wild} are by far the only publicly available FaaS workload traces that we have access to. The traces were collected during a two-week period, containing the average, minimum, maximum execution duration breakdown per function and invocation counts per function sampled at each one-minute interval. 
We analyzed the distribution of the average execution duration of all function invocations in the two-week period (Figure~\ref{fig:moti_azure_cdf}). We observe that the function execution duration spans a total of seven orders of magnitude; specifically, about $37.2\%$, $57.2\%$, and $99.9\%$ of the functions have an average execution duration shorter than 300~ms, 1~second, and 224~seconds, respectively. 

\noindent\textbf{Observation 1:}
\emph{While real-world FaaS workloads have a mixture of short and long functions, a majority of them are extremely short-lived and latency-sensitive. Optimizing the execution duration of these short functions will provide a huge benefit for the overall performance of FaaS platforms.}

\vspace{-4pt}
\subsection{OpenLambda Measurement}
\label{subsec:moti_ol}
\vspace{-2pt}

We next measured the performance of an Azure-sampled FaaS workload on 
OpenLambda. We generated the workload based on the Azure Functions workload datasets~\cite{serverless_in_the_wild}. 
Since our focus is on single-server scheduling,
we downscaled the original trace by sampling the execution duration and request inter-arrival times of $49,712$ function requests from Day 1. More details about workload generation are described in \cref{sec:methodology}.

We configured OpenLambda to use 12 CPU cores and Linux's real-time (RT) schedulers, {\small\texttt{SCHED\_FIFO}} (FIFO) and {\small\texttt{SCHED\_RR}} (RR), as well as Linux's default proportional-share scheduler, {\small\texttt{SCHED\_NORMAL}} (CFS). 
We tested the workload with two load levels, \added{an average load of  $80\%$ over all 12 cores, and an average of $100\%$ load,} and compared OpenLambda's performance against an offline oracle scheduler SRTF.
\added{SRTF always selects the job with the smallest remaining time to execute.}
SRTF is optimal as it assumes \emph{a priori} knowledge of function duration. 
\added{IDEAL scheduling represents the ideal scenario where there are infinite resources with zero contention.}

Figure~\ref{fig:moti} shows the performance and RTE results. 
In calculating the RTE, the aggregate CPU time of a function is measured under the IDEAL scenario while the turnaround time of the function is measured under the workload.
From this figure, we have the following observations.
(1)~SRTF, as an offline scheduling policy in favor of short jobs, is provably optimal for turnaround time~\cite{coffman1968computer}; 
SRTF approached the IDEAL performance, which was achieved with infinite resources.
(2)~None of Linux's three CPU schedulers was able to offer good performance under both the $80\%$ and $100\%$ load under practical FaaS workloads (Figure~\ref{fig:moti_perf}); 
CFS performed the best among all Linux scheduling policies, but still, there were about $11.4\%$ and $89.9\%$ of the function requests that achieved an RTE
score $<0.2$ (Figure~\ref{fig:moti_rte}). This explains why SRTF outperformed CFS: with the same service time, functions were preempted more under CFS, causing longer waiting times. 
(3)~Under the $100\%$ load, functions executed more than one order of magnitude slower under CFS than SRTF, with a $40^{th}$ and $70^{th}$ percentile slowdown of $16\times$ and $24\times$, respectively, again, because of the dominant waiting time.
(4)~RT schedulers offered the worst performance: FIFO performed the worst due to the ``convoy effect'', where short functions were blocked behind long functions.

\noindent\textbf{Observation 2:}
\added{\emph{Approximating the offline oracle SRTF by improving the run-time effectiveness will promise a significant performance boost for short serverless functions.}}

%% file: design.tex
\vspace{-2pt}
\section{{\proj} Design}
\label{sec:design}
\vspace{-2pt}

Our study in \cref{sec:motivation} shows that cloud functions often suffer high execution duration amplifications. 
Among all scheduling strategies, SRTF is promising (than CFS). This motivates the design of a new scheduler {\proj} to prioritize short functions. In this section, we present the design principle and challenges of {\proj}, followed by the design details.
\vspace{-6pt}
\subsection{Design Goals and Challenges}
\label{subsec:goals}
\vspace{-2pt}

To prioritize short functions, a priority-based scheduler is needed. However, as described earlier in \cref{subsec:sched_overview}, proportional-share schedulers such as CFS are designed for optimizing  fairness for long-running jobs and avoiding starvation. 
To achieve such goals 
when multiple concurrently running jobs are consolidated on a single server, CFS squeezes the time slice for each competing job and proportionally shares the physical CPU time among them. This leads to a significantly prolonged ``scheduling cycle'': a job that has used up its time slice is descheduled and must wait for a long time before it gets rescheduled. 
For short jobs, this prolonged waiting time hurts turnaround time: they could have finished much earlier without preemption
if given a \emph{long-enough} time slice.

Our motivational study from Figure~\ref{fig:moti} shows that SRTF can achieve much better performance than CFS.
SRTF provides a theoretical lower bound in terms of turnaround time for short-function-dominant FaaS workloads because SRTF allows a short-enough function to be scheduled instantly and run to completion without preemption. However, SRTF is not practical as it assumes \emph{a priori} knowledge about job duration. 

Our goal is to design an online scheduler that approximates SRTF. We achieve this goal by addressing the following challenges.
First, cloud functions are much shorter with a typical duration ranging from tens of ms to several seconds, and FaaS workloads exhibit transient overload.  
Such workload characteristics pose a challenge in designing effective prioritization strategies, which should prioritize short functions in a timely manner with minimal impact on longer functions.

Second, existing FaaS platforms use a client-server-based microservice architecture, where clients issue HTTP invocation requests to execute cloud function instances hosted by backend FaaS servers. Our design must provide a transparent and portable function scheduler that requires no or minimum modification of existing FaaS platforms while being OS-scheduler-agnostic. That is, even if the FaaS platform uses CFS or other OS-level proportional scheduling schemes, short functions should gain higher priority than longer functions. 
\added{Thus, a second challenge is how to design an efficient and practical function scheduler, which 
(1)~works entirely in the user space and does not require kernel modifications, 
(2)~works alongside (rather than replacing) OS schedulers and exploits (whenever needed) OS scheduling properties such as work conservation to provide better support for FaaS workloads, 
(3)~while being transparent to FaaS servers.}

\vspace{-4pt}
\subsection{Design Overview}
\label{design overview}
\vspace{-2pt}

\begin{figure}[t]
\begin{center}
\includegraphics[width=0.42\textwidth]{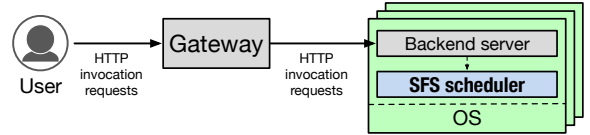}
\caption{Overview of a typical FaaS platform deployment.
}
\label{fig:global}
\vspace{-20pt}
\end{center}
\end{figure}

Typically, a FaaS platform uses a client-server architecture as depicted in Figure~\ref{fig:global}: a directly-user-facing gateway 
forwards HTTP invocation requests from users to a backend FaaS server that hosts requested function instances. 
To be transparent and portable to existing FaaS platforms, 
we design {\proj} by following a black-box approach. 
A backend FaaS server dispatches function invocations to the underlying OS. {\proj} assumes the existence of such a backend FaaS server. {\proj} serves as a user-space middle layer between a FaaS server and the OS (Figure~\ref{fig:global}), 
intercepting function requests and performing function scheduling for the FaaS server. 

Figure~\ref{fig:arch} illustrates the overall function scheduling flow. 
\added{{\proj} orchestrates Linux's existing schedulers (FIFO and CFS) in the user space.
{\proj} adopts two-level scheduling that seamlessly combines
a FIFO-like scheduling policy based on Linux FIFO
at top level and a kernel-space scheduling policy offloaded to Linux CFS at bottom level.}
The top-level scheduler schedules function requests by the order in which they are enqueued in the global queue and filters out those longer functions that are not finished in a (dynamically configurable) time slice. This way, the top-level scheduler effectively serves as a 
{\bf FILTER}\footnote{FILTER: {\bf F}irst {\bf I}n but {\bf L}onger jobs {\bf T}o {\bf E}xtra {\bf R}unqueue.}. 
Under {\proj},  a function's lifespan may experience one or two phases: 
a \emph{FILTER phase} 
and/or a \emph{CFS phase}.
A function by default starts in FILTER mode. {\proj} dynamically adapts a \emph{time slice} parameter $S$ (discussed later) using a sliding window approach and uses $S$ to bound a function's execution in FILTER mode. This way, {\proj} approximates SRTF.
{\proj} is inherently work-conserving following a single queue model:
\added{{\proj} workers fetch requests whenever they are idle.} 
To minimize context switches, {\proj} guarantees that
a function that is executing in FILTER mode would be preempted only if it has used up the time slice or it is waiting for an event (e.g., I/O). 

\noindent\textbf{Scheduling Flow.}
Next, we describe the main components and the scheduling flow of {\proj} as illustrated in Figure~\ref{fig:arch}: 

\begin{enumerate}[noitemsep,leftmargin=*]

\item[1.] A backend FaaS server, serving as a client, dispatches function invocation requests, launches requested functions in a virtualization environment in OS, and sends the information of the dispatched function requests
(tuples of unique function request ID and the invocation timestamp) to {\proj}' {\bf global queue}. 

\begin{figure}[t]
\begin{center}
\includegraphics[width=0.48\textwidth]{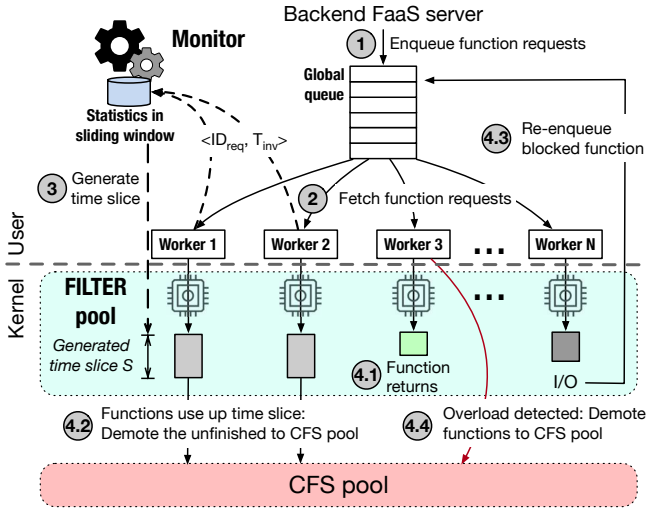}
\caption{{\proj} architecture. {\proj}' components are highlighted in condensed bold font. 
}
\label{fig:arch}
\vspace{-20pt}
\end{center}
\end{figure}

\item[2.] 
Multiple {\proj} \textbf{workers}, each responsible for scheduling function requests on a separate CPU core, 
concurrently fetch function requests from the global queue whenever workers are free.
Note a non-empty global queue indicates that all cores are busy serving requests. Each {\proj} worker is responsible for intercepting the dispatched function process by using the tuple information fetched from the queue and scheduling the function using FILTER policy. 
This way, each function by default starts execution in FILTER mode, unless otherwise specified (\cref{subsec:overload}).
This effectively forms a \textbf{FILTER pool} of multiple {\proj} workers. 
\added{Functions that are executing under FILTER mode naturally gain higher priority than those under CFS mode. To realize this, {\proj} changes a running function process' OS scheduling policy to FIFO ({\small\texttt{SCHED\_FIFO}}), which has higher static priorities than CFS ({\small\texttt{SCHED\_NORMAL}}) processes~\cite{linux_sched}.}

\item[3.] Each {\proj} worker stores the following statistics information in memory:
(1)~function request ID and its invocation timestamp, which is initially recorded in the global queue when the function request was submitted,
and (2)~function execution timestamp, which is the time when the function starts execution. 
An {\proj} \textbf{monitor} periodically re-calculates a global \emph{time slice} parameter $S$ based on the collected statistics (\cref{subsec:timeslice}). 
Next, we describe several cases of the function execution. 

\item[4.1.] The ideal case is that a short function finishes execution and returns before using up $S$. This way, the {\proj} worker marks its completion, removes the corresponding entry from the global queue, fetches the next function request, and restarts the time slice timer. 

\item[4.2.] The {\proj} worker keeps track of the runtime of the FILTER function, forcibly preempts it if its time slice expires, and demotes it to CFS.

\item[4.3.] If a function is blocked by some event, e.g., an I/O event, the worker will instantly preempt its execution and add it back to the global queue (\cref{subsec:io}). 

\item[4.4.] If an {\proj} worker detects a transient overload by observing increasing queuing delay (above a certain threshold), it temporarily disables FILTER and directly schedules next requests using CFS (\cref{subsec:overload}). 

\end{enumerate}

\vspace{-4pt}
\subsection{Dynamically Adapting Time Slices}
\label{subsec:timeslice}
\vspace{-2pt}

While it may be impossible to design a perfect scheduler that assigns a precise time slice that perfectly matches the remaining execution duration of any job, {\proj} uses a simple yet effective heuristic approach based on queuing theory to estimate and dynamically adapt the time slice parameter $S$. 

The time slice parameter 
presents an interesting tradeoff between queuing delays and the turnaround time of the workload. On the one hand, an (improperly) short time slice value would reduce global queuing delays of outstanding function requests but unnecessarily cause a longer turnaround time due to increased context switches. CFS falls to this end of the spectrum.
On the other hand, an (unnecessarily) long time slice value would lead to reduced overall context switches but increased global queuing delays, and as a result, hurts turnaround time. 

Therefore, we use queuing theory to set the time slice.
For this purpose, we can model a multi-core scheduler as a multi-server queuing system---an $M/G/c$ model according to Kendall’s notation---using the following equation:
\vspace{-4pt}
\begin{equation} \label{eq:traffic}
  \rho = \frac{\lambda}{c\mu} 
\vspace{-3pt}
\end{equation}
where $\lambda$ is the arrival rate of the requests, $\mu$ is the service rate of a single core, $c$ is the number of cores used, and $\rho$ is the traffic intensity per core (i.e., per-core utilization). We can thus use the utilization metric $\rho$ as a measure of queuing delay: 
if $\rho$ is greater than one, meaning the arrival rate $\lambda$ is larger than the aggregate service rate $c\mu$, the length of {\proj}' global queue will grow without bound. 
Intuitively, adapting the service rate $\mu$ based on the changing $\lambda$ can bound $\rho$, thus the overall queuing delay. 
However, in practice, $\mu$ is solely determined by the workload and the capacity of the underlying hardware. Therefore, {\proj} enforces a global time slice to cap the duration for how long any function may run in FILTER mode. {\proj} dynamically changes the time slice in response to the variable request arrival rate, which is estimated using historical IATs.

Following Equation~\ref{eq:traffic}, {\proj} keeps track of a small sliding window of last $N$ requests' inter-arrival times (IATs) to determine the time slice parameter $S$.
For a single-core system, {\proj} calculates the average IAT of last $N$ functions, $\overline{IAT}$, and uses it as the feedback to dynamically tune $S$. 
With $c$ cores, $S = \overline{IAT} * c$. 
Intuitively, $S$ is used to bound the service rate $\mu$ of Equation~\ref{eq:traffic}, which in turn affects the traffic intensity $\rho$; $\rho$ further affects the queuing delay of function requests that are executing under FILTER mode, and therefore, {\proj} uses the historical IAT information to strike a balance between queuing delay and execution time.
When a global $S$ is selected, {\proj} guarantees that all functions whose execution duration is shorter than $S$ run to completion without being preempted. {\proj} re-calculates a new $S$ for every $N$ function requests that has been enqueued in order to provide dynamic adaptation to workloads. $N$ is configurable and we choose 100 as $N$ in our evaluation.

There may always be functions that are not able to finish before the time slice elapses. To solve this issue, {\proj} uses a single-level FILTER pool concatenated with CFS to approximate SRTF. 
\added{{\proj} 
steers Linux FIFO directly from the user space and builds the FILTER policy as a high-priority queue for short functions.}
{\proj} transparently leverages CFS as a black-box, lower-priority queue for longer functions. Note that functions running in CFS share the same set of cores as those running in FILTER mode. 
Starvation is mitigated since CFS is work-conserving and can immediately schedule any demoted functions on any available CPU core.

\vspace{-4pt}
\subsection{Handling I/Os}
\label{subsec:io}
\vspace{-2pt}

Since {\proj} is a user-space scheduler, it cannot transparently handle kernel-level tasks such as context switches, interrupts, and preemptions, etc. 
That is, an {\proj} worker could be waiting for a blocked function that has already been preempted due to an I/O event. This leads to sub-optimal decision-making with regard to function timekeeping and time slice estimation. To solve this issue, {\proj} workers track the kernel-level process status of the function by periodically issuing a polling request to the OS. When a function is in its CPU burst, its kernel-level status is in {\small\texttt{running}} mode. Whenever a function changes its kernel-level status from {\small\texttt{running}} to {\small\texttt{sleep}},
the {\proj} worker detects this transition, stops its timekeeping and records the unused time slice, reduces its priority, and schedules the next available function from the global queue. Note that, when a high-priority function blocks by I/O, CFS automatically sneaks in and executes other functions that have been filtered by {\proj}. This guarantees seamless work conservation.
When the status of a waiting function changes to {\small\texttt{runnable}}, {\proj} adds it back to global queue. When this function gets rescheduled in the FILTER pool, it will execute until it completes or it uses up the rest of the time slice. We use 4~ms as the polling interval. We evaluate this scheme in \cref{subsec:sensitivity_analysis} and its overhead in \cref{subsec:ol_overhead}.

\vspace{-4pt}
\subsection{Handling Overload}
\label{subsec:overload}
\vspace{-2pt}

Real-world FaaS workloads exhibit highly bursty and unpredictable load patterns~\cite{serverless_in_the_wild, faasnet_atc21}. Alibaba Function Compute workload analysis reports transient spikes of concurrent invocations to the same function~\cite{faasnet_atc21}.  
When an increasing number of short functions get piled up at global queue in a very short time (increasing arrival rate $\lambda$ in Equation~\ref{eq:traffic}), the service rate of {\proj}' FILTER pool, $c\mu$, cannot catch up with the workload spike. This transient (temporary) overload leads to increased traffic intensity, $\rho$, therefore, increased queuing delay and even function request drop. 
Reducing the time slice of the FILTER pool helps little in this scenario. 
This is because a FILTER time slice shorter than that of CFS would cause more context switches than CFS; as a result, the piled-up FIFO function requests from the transient overload  create backlog that cannot be quickly consumed by FILTER workers (see Figure~\ref{fig:delay} as an example).
To solve this issue, {\proj} temporarily switches to CFS when any {\proj} worker detects increasing queuing delay of the function request that it is about to schedule using FILTER. As long as the queuing delay lowers back to normal, {\proj} workers roll back to the normal scheduling flow.

This strategy, though simple, is in fact very effective because of the following reasons. 
Offloading accumulated functions to CFS alleviates high queuing delays in the FILTER pool by draining the backlog more quickly. 
Since overload is transient, regular load coming after that can then be serviced by {\proj}' default, time-slice-based FILTER pool first, thus, short functions experience no further queuing delays and can finish in one round before the time slice expires. 
Those function requests that are offloaded during the overload to CFS are eventually complete thanks to CFS' work conservation. 
An {\proj} worker detects overload if the queuing delay is at least $O\times S$. We set $O$ as 3 empirically.
We evaluate the efficacy of this strategy in \cref{subsec:sensitivity_analysis}. 

%% file: implementation.tex
\section{{\proj} Implementation}
\label{sec:impl}
\vspace{-2pt}

We have implemented {\proj} as a standalone, user-space function scheduler in Go.
We have also ported {\proj} to an open-source FaaS platform OpenLambda~\cite{openlambda_hotcloud16}. 
Porting  {\proj} to OpenLambda required a very small engineering effort: we modified/added 29 lines of Go/Python code in OpenLambda worker and sandbox server to interface with {\proj}.

\noindent\textbf{Standalone {\proj}.}
We implemented the {\proj} global queue structure using Go's built-in, thread-safe channel.
{\proj} workers are goroutines (a lightweight user-level thread of execution managed by the Go runtime), which are responsible for dispatching and scheduling function processes in the FILTER pool. 
Function invocation requests are pushed into the global queue channel by 
an external, backend FaaS server  (Figure~\ref{fig:global}-\ref{fig:arch}). {\proj} implements the switching from FILTER pool to CFS pool by using Linux {\small\texttt{schedtool}}~\cite{schedtool}. Function invocation requests are executed in sandboxed processes scheduled by either {\proj} workers or the CFS scheduler. 

We chose to implement a global queue instead of using a per-core-queue (i.e., multi-queue) design because a single global queue guarantees natural work conservation with good load balancing across all CPU cores. It is demonstrated that a per-core-queue design has multiple downsides, e.g., severe load imbalance, core under-utilization, and degraded performance~\cite{prekas2017zygos}. In our current implementation, the global queue is implemented using a Go channel (with nanosecond enqueue/dequeue latency under multi-threaded environments), which is capable of handling up to 100 CPU cores each running tasks with a duration from ms to seconds. A single global queue, however, might become a bottleneck if assuming hundreds of CPU cores and microsecond-level function execution duration. Since the server hosts deployed by a typical FaaS provider may have up to 100 vCPUs~\cite{firecracker_nsdi20}, our global queue design is an appropriate solution.

{\proj} workers are work-conserving: each worker is blocked on global queue and fetches a function request whenever the queue has entries to consume. To implement FILTER policy, {\proj} workers use {\small\texttt{schedtool}} to change a running function process' OS scheduling policy from CFS ({\small\texttt{SCHED\_NORMAL}}) to FIFO ({\small\texttt{SCHED\_FIFO}}). When the server is under low utilization, a function that is dispatched by the FaaS server may execute in CFS for a very short period of time (hundreds of microseconds, depending on the communication overhead between the FaaS server and {\proj}). Under this situation, CFS performs the same as {\proj} due to zero contention.
A non-empty global queue indicates that all CPU cores are busy serving function requests; if so, newly dispatched functions are internally queued at OS-level run queues (\cref{subsec:sched_overview}) as CFS jobs have inherently lower priority than actively running FIFO jobs. 
{\proj} workers preempt a FIFO function's execution by using {\small\texttt{schedtool}} to assign it a lower priority. 
{\proj} workers detect blocking events by periodically (4~ms) polling the function process' OS status using \added{a go process utilities library {\small\texttt{gopsutil}}}~\cite{gopsutil}.

\begin{figure}[t]
\begin{center}
\includegraphics[width=0.45\textwidth]{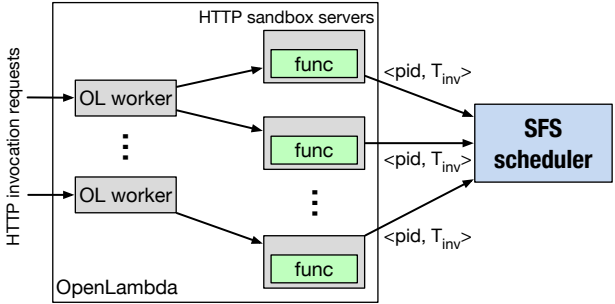}
\caption{Porting {\proj} to OpenLambda. OL: OpenLambda.}
\label{fig:impl}
\vspace{-20pt}
\end{center}
\end{figure}

{\proj} is designed to be, in principle, portable to any open source FaaS platforms such as OpenLambda~\cite{openlambda_hotcloud16}, OpenWhisk~\cite{openwhisk}, and OpenFaaS~\cite{openfaas}, as they all share the same client-server architecture. To demonstrate the portability, we have ported {\proj} to OpenLambda. 
As shown in Figure~\ref{fig:impl}, a backend OpenLambda deployment consists of two components: 
(1)~a cluster of OpenLambda workers that are responsible for receiving function invocation requests, performing sandbox auto-scaling, and tracking statistics,
and (2)~a cluster of HTTP servers that manage function sandboxes. In our implementation, we chose Docker containers as the function sandbox. On-demand container provisioning (i.e., cold start) incurs high overhead. Therefore, we disabled auto-scaling and pre-warmed enough function containers to simulate a stable-phase FaaS backend so as to accurately quantify the performance of schedulers.

\noindent\textbf{Porting {\proj} to OpenLambda.}
We modified OpenLambda's HTTP sandbox server to communicate with {\proj} scheduler using UDP: whenever a sandbox server dispatches a function request to OS, it sends to {\proj} a UDP message containing function process PID and invocation timestamp.

%% file: evaluation.tex
\vspace{-2pt}
\section{Experimental Methodology}
\label{sec:methodology}
\vspace{-2pt}

\noindent\textbf{Setup.}
We developed and tested {\proj} on CloudLab~\cite{cloudlab_atc19}.
We evaluated {\proj} and {\proj}-ported OpenLambda on two AWS EC2 VMs: a small, {\small\texttt{c5a.4xlarge}} VM with 16~vCPUs and 64~GB memory (standalone {\proj}), and a large, bare-metal, {\small\texttt{m5.metal}} EC2 instance with 96~vCPUs and 384~GB memory (OpenLambda).  The goal of using a large bare-metal machine is to simulate a similar FaaS deployment environment used by major FaaS providers such as AWS Lambda~\cite{firecracker_nsdi20}.

\noindent\textbf{{\bench}.}
We have built a FaaS workload generator called {\bench}, which creates FaaS workloads modeled after the Azure Functions workload~\cite{serverless_in_the_wild}. {\bench} is highly configurable along the following dimensions:
(1)~{\bench} configures per-function behaviors by using a Fibonacci ({\fib}) function with two knobs: an integer knob $N$ for controlling the compute time, and a boolean knob $IO$ that toggles the I/O operation (if set true)
to simulate I/O-intensive functions. 
The distributions of (2) function duration and (3) requests' inter-arrival times (IATs) are also configurable.

\begin{figure*}[t]
\begin{minipage}{\textwidth}
\begin{center}
\begin{minipage}[b]{0.37\textwidth}
\includegraphics[width=1\textwidth]{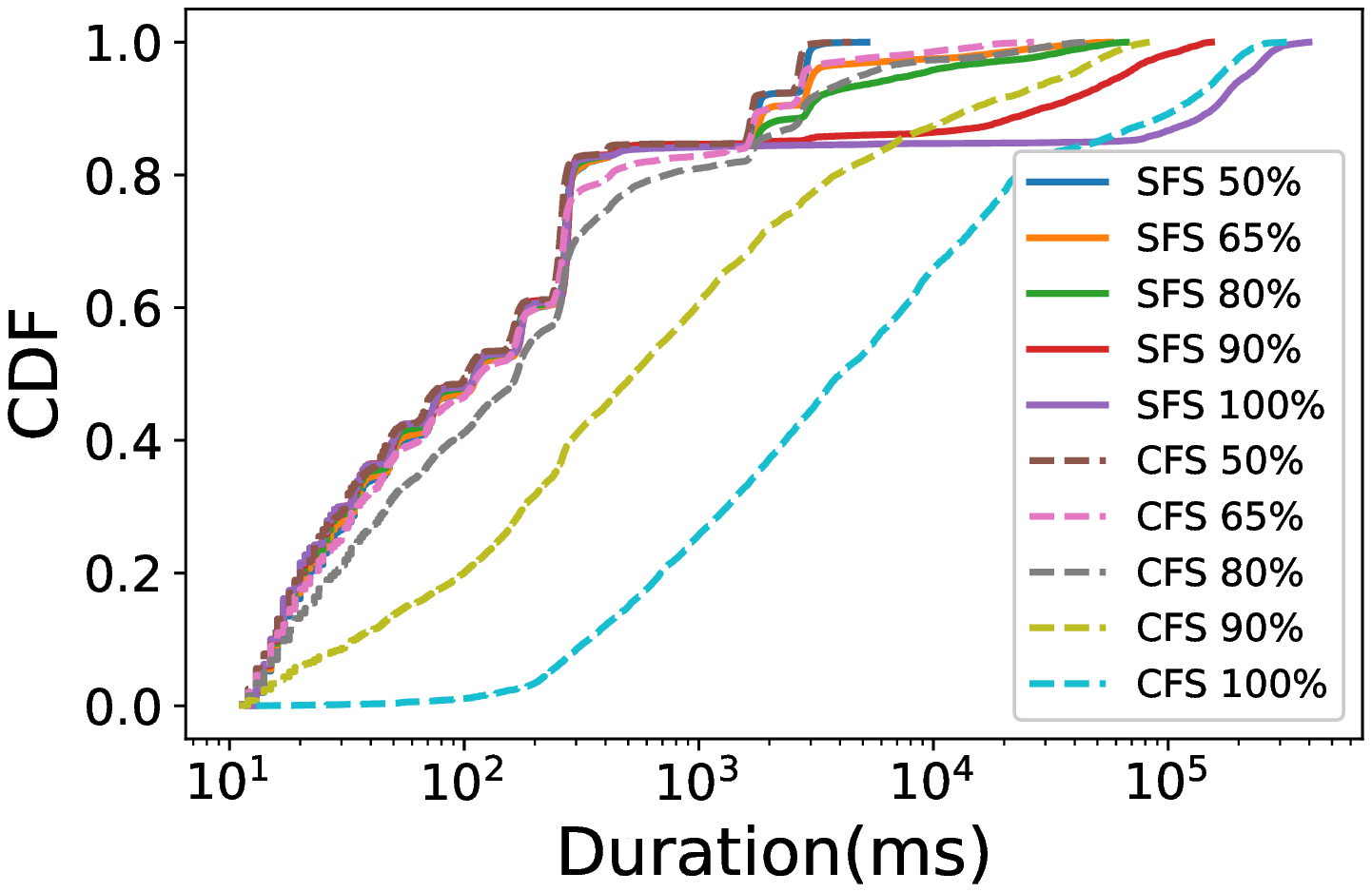}
\vspace{-15pt}
\caption{
Performance CDF.
}
\label{fig:load_perf}
\end{minipage}
\hspace{30pt}
\begin{minipage}[b]{0.37\textwidth}
\includegraphics[width=1\textwidth]{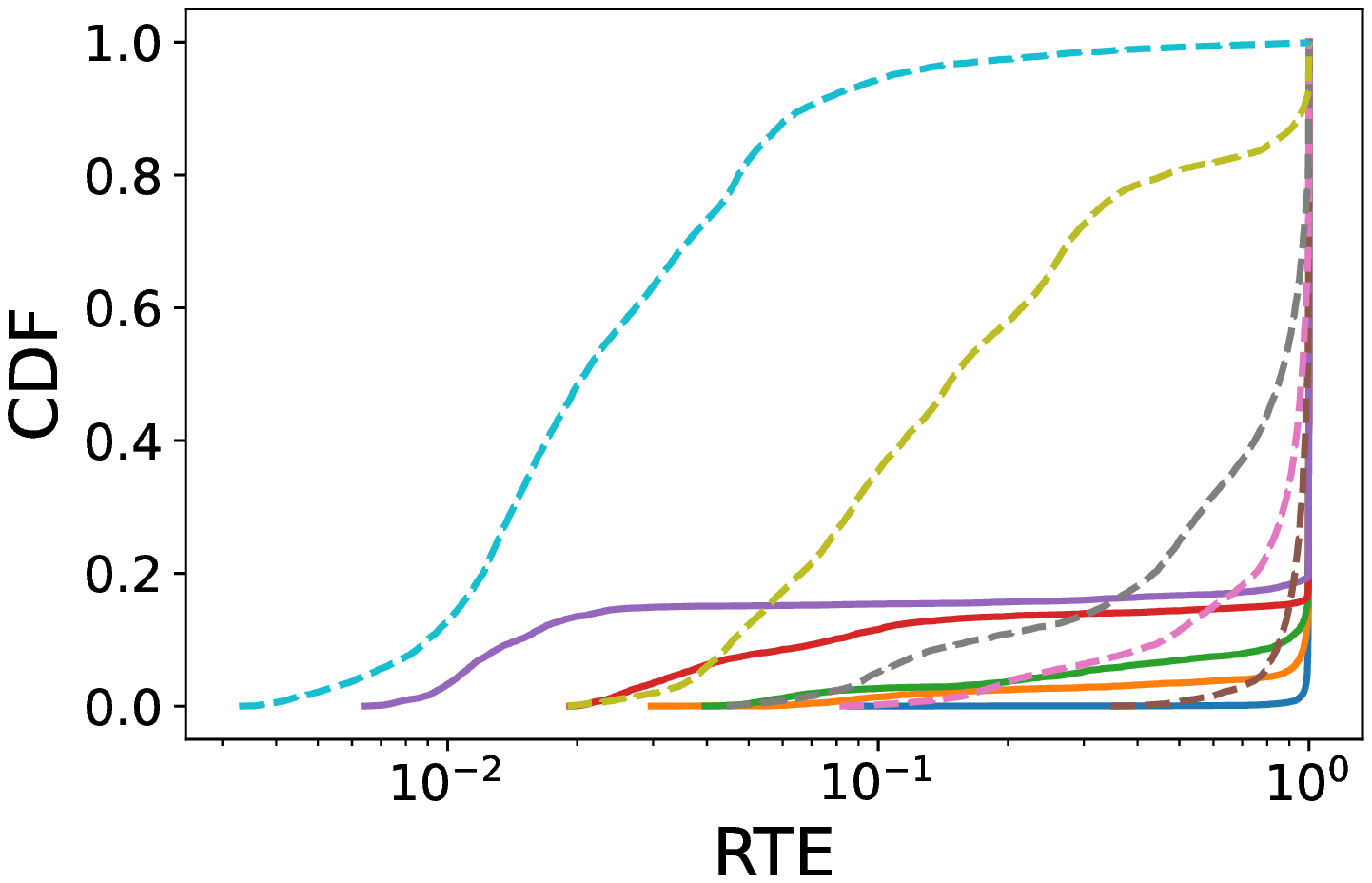}
\vspace{-15pt}
\caption{
RTE CDF. }
\label{fig:load_fre}
\end{minipage}
\end{center}
\end{minipage}
\vspace{-10pt}
\end{figure*}

\begin{figure}
\begin{center}
\includegraphics[width=0.47\textwidth]{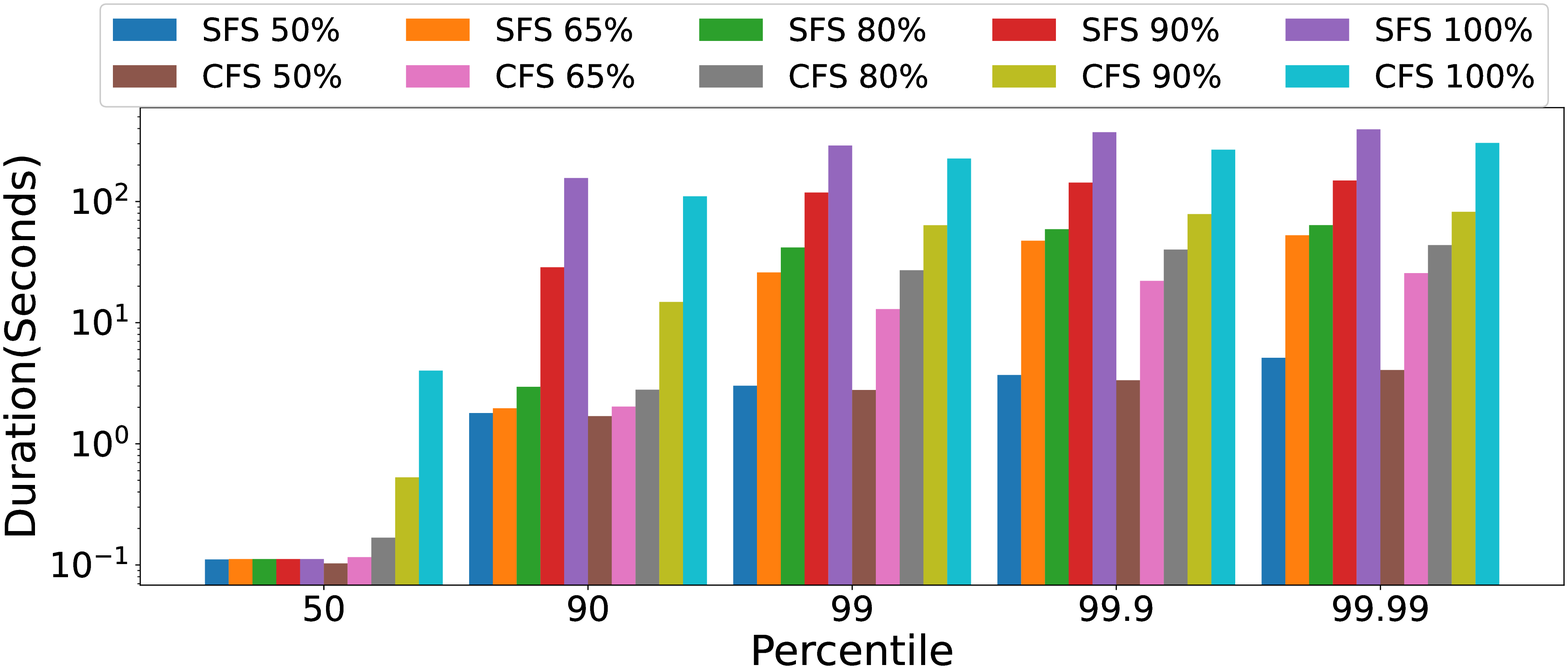}
\vspace{-5pt}
\caption{
Percentile breakdowns of function execution duration.
}
\vspace{-20pt}
\label{fig:standalone_percentile}
\end{center}
\end{figure}

\begin{wraptable}{r}{0.62\columnwidth}
\vspace{-10pt}
\centering
\caption{
Probability distribution of function duration ranges and the corresponding {\fib} $N$s. Note ranges are non-contiguous and each missing range has less than $1\%$ probability in Azure traces (Day 1). 
}
\scalebox{0.95}{
\begin{tabular}{lrr}
\Xhline{2\arrayrulewidth}
{\bf Probability} & {\bf Duration} & {$N$} \\
\hline
{$40.6\%$} & {0-50~ms}  & {20-26} \\
{$9.8\%$} & {50-100~ms} & {27-28} \\
{$6.8\%$} & {100-200~ms} & {29} \\
{$22.7\%$} & {200-400~ms} & {30-31} \\
{$15.7\%$} & {$\geq 1550$~ms} & {34-35} \\
\Xhline{2\arrayrulewidth}
\end{tabular}
} 
\label{tbl:map}
\vspace{-6pt}
\end{wraptable}
\noindent\textbf{Generating FaaS Workloads.}
We based the Azure Functions traces to generate testing workloads. The original Azure traces contain the execution duration, memory sizes, and invocation timestamps of $82,375$ unique function applications spanning a period of 14 days. To downscale, we  generated the distribution of function execution duration based on Day 1's invocation statistics. 
We found that the duration roughly follows a multimodal distribution, where about $40.6\%$, $22.7\%$, and $15.7\%$ of invocations fall in a duration (unit of ms) range of $(0, 50]$, $[200, 400)$, and $[1550, \infty)$, respectively. We built a table that maps the range of execution duration recorded in Day 1 to {\fib}'s $N$s (Table~\ref{tbl:map}), and then used Azure function duration distribution to generate $N$s using {\bench}.
For example, {\fib} with an $N$ between 20-26 finishes execution in less than 45~ms, therefore, we programmed {\bench} to generate {\fib} functions with an $N$ between 20-26 with a probability of $40.6\%$. 
{\bench} can also generate different duration distributions (results omitted due to page limit).

The released Azure trace datasets only contain the high-level statistical breakdown information that describes the function execution behaviors. To make sure that our study captures the original Azure Functions workload as accurately as possible, we took the $50^{th}$ percentile execution duration as the expected execution time for a function. This way, our benchmark rules out outliers that do not represent the typical behaviors of the original workload.

To configure IATs of the generated workload, we randomly sampled 100 unique function applications, each with a total invocation count greater than 200 on Day 1, and extracted the IAT statistics. We then replayed the first $10,000$ invocation requests by strictly following the extracted IAT patterns. This is to guarantee that our generated workload preserves similar load patterns as real-world, production workloads. \added{In addition to modeling existing trace's IATs, {\bench} can also generate Poisson and uniform IATs.} 
We ran each test multiple times and results had negligible variation across runs.

\vspace{6pt}
\noindent\textbf{Goals.} Our evaluation aims to answer the following questions:
\vspace{-8pt}
\begin{itemize}[noitemsep,leftmargin=*]
    \item How does {\proj} perform under various loads (\cref{subsec:load})? 
    \item How do different {\proj} configurations affect its performance (\cref{subsec:sensitivity_analysis})? 
    \item How does an {\proj}-ported FaaS platform (OpenLambda) perform (\cref{sec:ol})?
\end{itemize}
\vspace{-2pt}

\vspace{-2pt}
\section{Evaluating Standalone {\proj}}
\label{sec:standalone}
\vspace{-2pt}

In this section, we evaluate {\proj} as a standalone function scheduler using {\bench}. 
The goal of evaluating standalone {\proj} is to better understand the true performance characteristics of task scheduling without the extra overhead introduced by a FaaS platform.

\vspace{-2pt}
\subsection{SFS Efficiency under Various Loads}
\label{subsec:load}
\vspace{-2pt}

We first test {\proj} under different loads using the Azure-sampled workload generated by {\bench}, which follows the duration distribution of Azure traces (Table~\ref{tbl:map}) with a Poisson IAT distribution. We adjusted the IAT of the generated workload proportionally to simulate different loads ranging from $50\%$ to $100\%$ of overall CPU utilization across all CPU cores. Figure~\ref{fig:load_perf} reports the CDF of the execution duration. {\proj} performed almost the same as CFS under the lowest $50\%$ load and slightly outperformed CFS under medium loads when the load increased from $65\%$ to $80\%$. {\proj}' marginal improvement was obtained because of a higher RTE. As shown in Figure~\ref{fig:load_fre}, about $93\%$ and $88\%$ of all function requests receive an RTE  $\geq 0.95$ under a load of $65\%$ and $80\%$, respectively, indicating that these functions run to completion without any context switch under {\proj} (with a very short queuing delay when the request was initially submitted). CFS is workload-oblivious, which introduces more context switches; under a load of $65\%$ and $80\%$ with CFS, only $55\%$ and $35\%$ of all function requests receive an RTE score $\geq 0.95$, 
where a lower RTE translates to prolonged waiting time. 

\added{An interesting observation 
is that {\proj} maintains almost identical performance for $83\%$ of the function requests across all load levels. In other words, at least $83\%$ of the function requests can achieve optimal execution duration and an RTE score of almost 1 even under a high load where all CPUs are $100\%$ utilized.

Under CFS, the same set of functions, on the other hand, saw dramatically increased execution duration because of 
prolonged waiting time (Figure~\ref{fig:load_fre}). This result demonstrates {\proj}' efficacy in sustaining dynamic FaaS workloads.} 

\added{The performance gain of shorter functions under {\proj} does not come for free: there is always a tradeoff in balancing the scheduler efficiency for short and long jobs~\cite{light_tail_or12}. {\proj} observed slightly higher tail latency. For the $17\%$ relatively long functions, {\proj} observed an average increase of $1.29\times$ in execution duration compared to CFS under the $100\%$ load. The $99.9^{th}$ percentile latency of {\proj} under $80\%$ load is only $47.1\%$ higher than that of CFS (Figure~\ref{fig:standalone_percentile}). CFS, while being a proportional-share scheduler, does suffer long tail latency even under relatively lower load; this can be seen from the fact that the $99.9^{th}$ percentile latency of CFS increases from 3.3~seconds under the $50\%$ load to 22.1~seconds under the $65\%$ load; though the increase of the $99.9^{th}$ percentile tail latency  under {\proj} is slightly higher than that under CFS.

Interestingly, {\proj} achieved a consistent medium ($50^{th}$ percentile) latency of 0.1~second across all load levels, while CFS' medium latency increases as the load increases.} 
Longer functions could be potentially offloaded to relatively lighter-loaded FaaS servers by the global FaaS scheduler to mitigate the performance impact, which we plan to investigate as part of our future work.

\added{More importantly,
even under a $100\%$ load, {\proj} offers short functions consistently competitive performance comparable to a performance level that would have been achieved under less stringent load situations, say $65\%$--$90\%$ load.
This would bring desired benefits for both parties, including mitigated overcharges for FaaS users and higher resource utilization (thus reduced deployment costs) for FaaS providers.}

\begin{figure*}[h]
\begin{minipage}{\textwidth}
\begin{center}
\begin{minipage}[b]{0.3\textwidth}
\vspace{5pt}
\includegraphics[width=.99\textwidth]{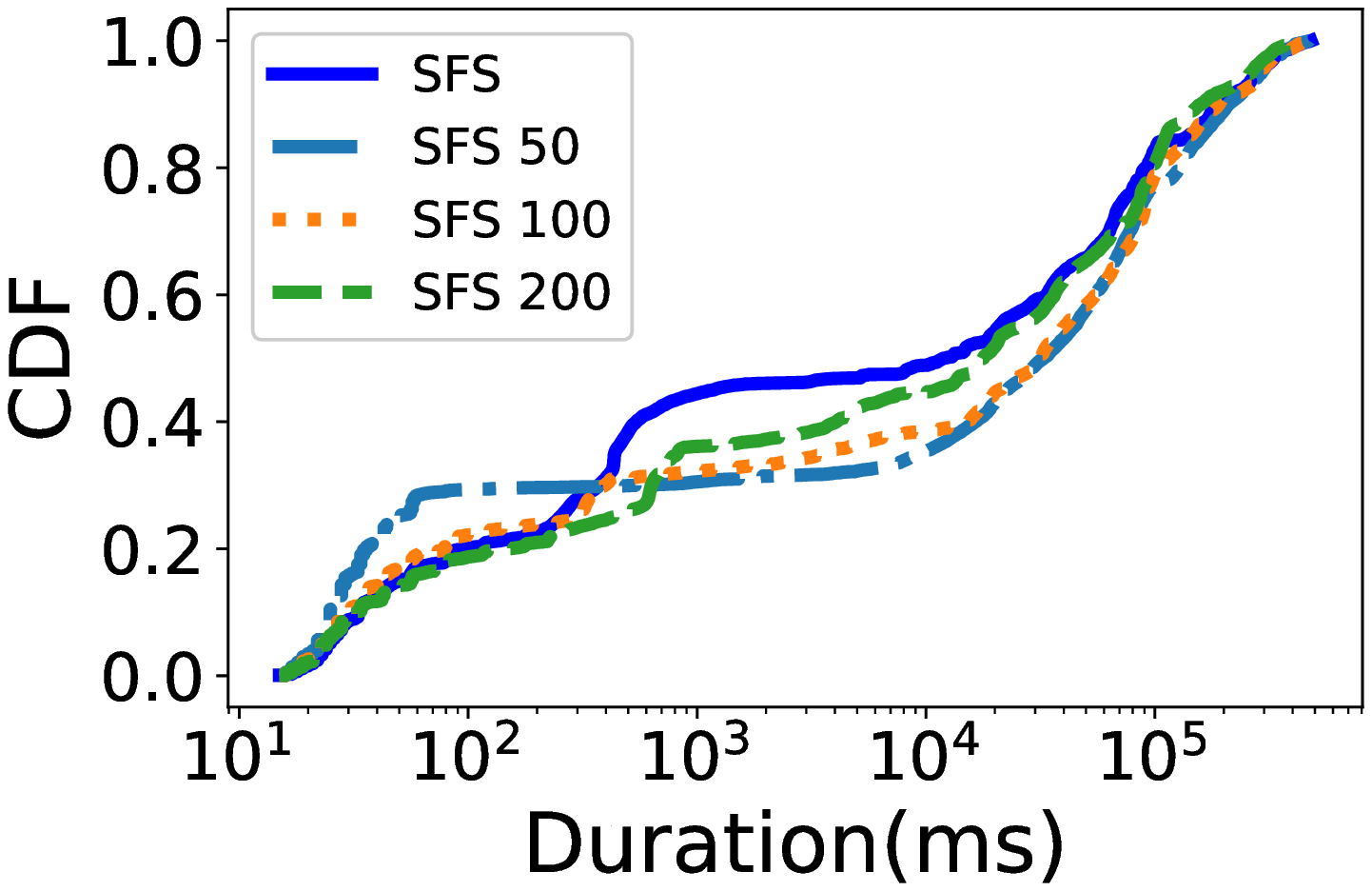}
\caption{
Adaptive time slice tuning vs. statically fixed time slices.
} 
\label{fig:ts_cdf}
\vspace{-10pt}
\end{minipage}
\begin{minipage}[b]{0.345\textwidth}
\includegraphics[width=1\textwidth]{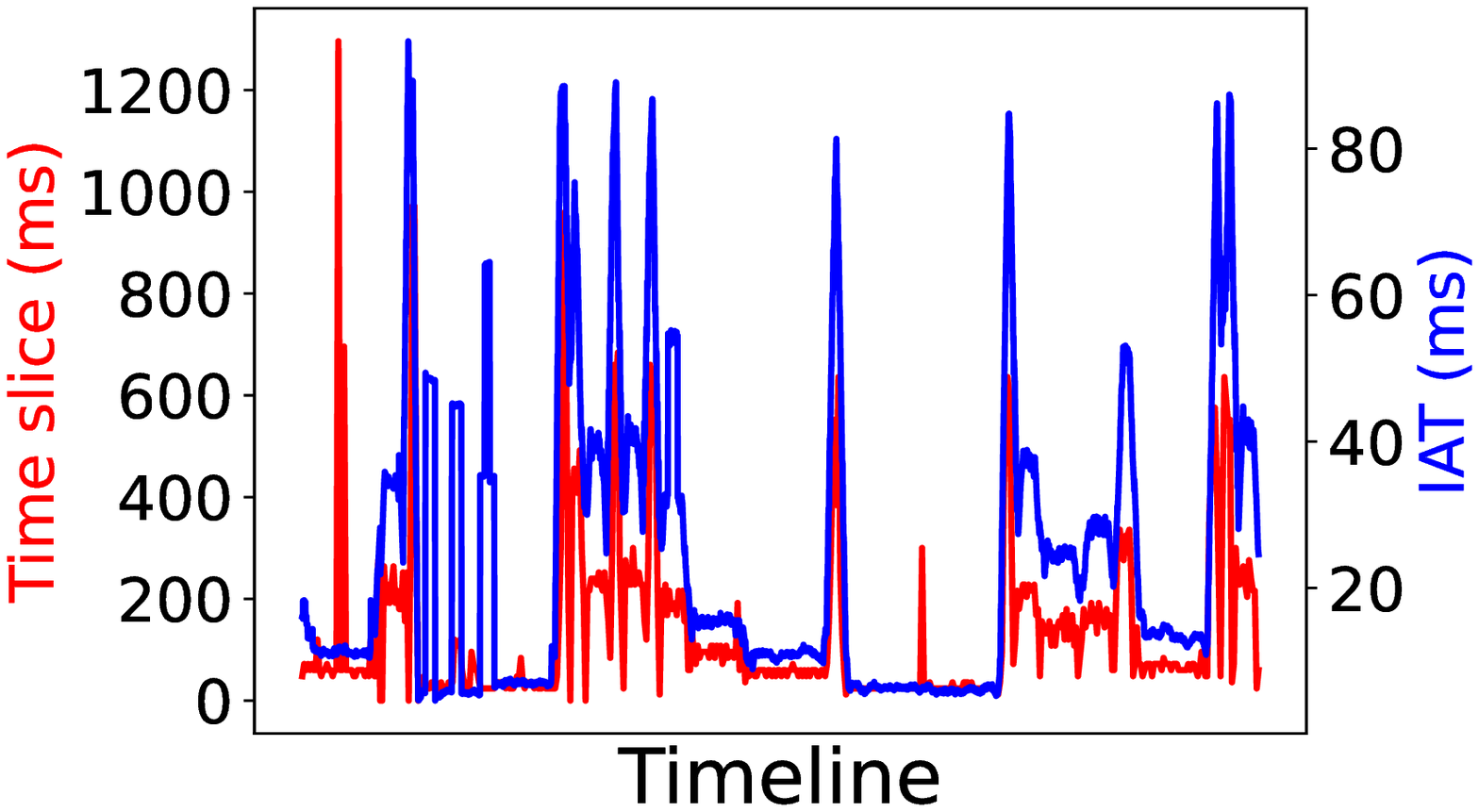}
\caption{
Timeline of time slice changes vs. IATs during the whole workload.
}
\label{fig:ts_t}
\vspace{-10pt}
\end{minipage}
\begin{minipage}[b]{0.3\textwidth}
\includegraphics[width=0.99\columnwidth]{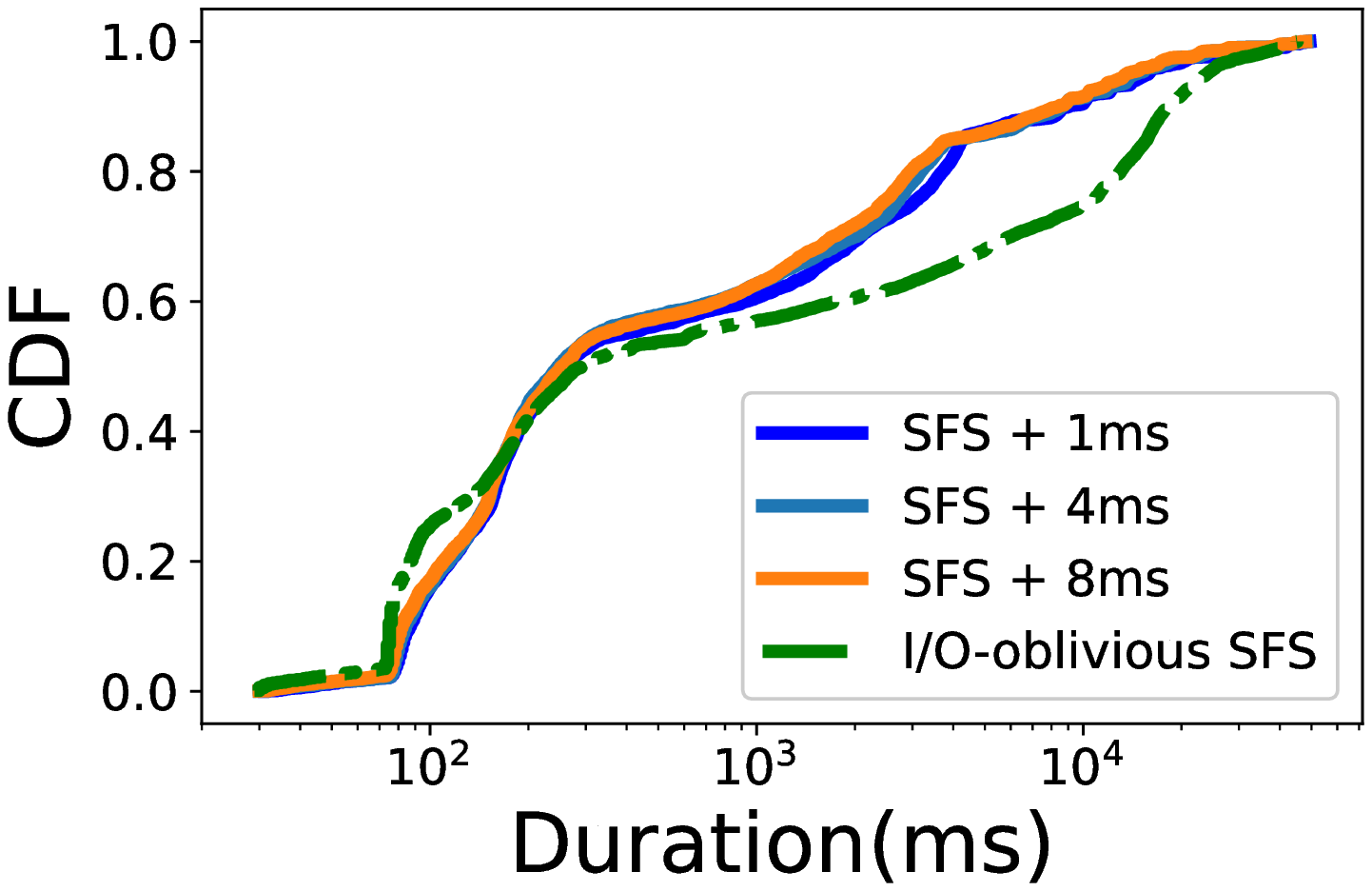}
\caption{Handling I/O. {\proj} was configured to use different polling intervals.
}
\label{fig:io}
\vspace{-10pt}
\end{minipage}
\end{center}
\end{minipage}
\end{figure*}

\vspace{-4pt}
\subsection{Sensitivity Analysis}
\label{subsec:sensitivity_analysis}
\vspace{-2pt}

\noindent\textbf{Impact of Time Slice Configurations.}
Next, we conduct a sensitivity test by varying the time slice parameters $S$. 
We fixed $S$ to 50, 100, 200 and compared against {\proj}'s dynamic adaptation heuristic.
Figure~\ref{fig:ts_cdf} shows that none of the three statically configured $S$ led to optimal performance. {\proj}'s adaptive strategy yields better performance than a static $S$ of 100~ms and 200~ms by adapting $S$ based on the last 100 observed IAT samples. This is because a long, fixed time slice inevitably increased queuing delay of waiting functions.

Figure~\ref{fig:ts_t} depicts the timeline of the adaptation.  
Having a smaller, fixed $S$ as short as 50~ms resulted in better performance for around $30\%$ short function requests compared to {\proj}, but at the same time, it suffered from significantly prolonged duration for the rest of $70\%$ requests. {\proj} struck a good balance of queuing delays and service time, leading to better overall performance.  

\noindent\textbf{Handling I/O.}
To evaluate 
how {\proj} handles I/O events in functions, 
we toggled the I/O knob for $75\%$ of the function requests, for which we added a single I/O operation at the beginning of the function execution; the added I/O operation took $X$ ms, where $X$ was randomized drawn from a range between 10 to 100~ms.
As shown in Figure~\ref{fig:io}, I/O-oblivious {\proj} had worse performance, because FILTER pool wasted time slice credit waiting for the I/O to be served, causing them to be filtered out to CFS. In contrast, {\proj} was able to detect I/O-caused waiting by using periodic status polling. 
We varied the polling interval from 1~ms to 8~ms and found that the performance was not sensitive to the polling frequency.

\begin{figure*}
\begin{center}
\hspace{10pt}
\subfigure[Timeline of queuing delays.] {
\includegraphics[width=.37\textwidth]{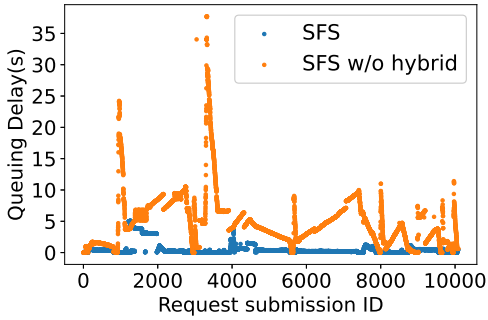}
\label{fig:delay}
}
\hspace{12pt}
\subfigure[CDF of function duration.] {
\includegraphics[width=.37\textwidth]{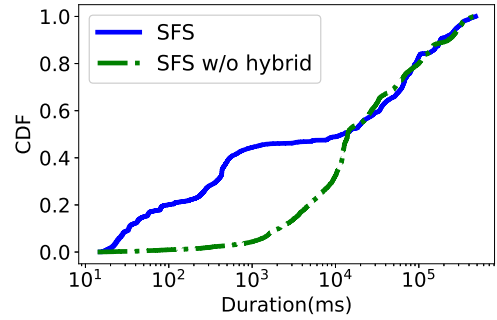}
\label{fig:switch_cdf}
}
\vspace{-5pt}
\caption{
Effect of {\proj}'s overload handling mechanism. 
{\small\texttt{SFS w/o hybrid}} refers to {\proj}'s baseline implementation with the hybrid FILTER+CFS mode disabled (see \cref{subsec:overload}).
}
\label{fig:switch}
\end{center}
\vspace{-10pt}
\end{figure*}

\noindent\textbf{Handling Overload.}
We finally test the effectiveness of {\proj}'s hybrid strategy to handle the transient overload. The Azure-sampled workload exhibits transient overloads, which can be spotted from the five queuing delay spikes shown in Figure~\ref{fig:delay}. Note we only measured the time a function spent waiting in {\proj}' global queue. 
With overload detection disabled, {\proj} suffered significantly long queuing delays. Spiked queuing delays took long to diminish because the normal workload coming after the temporary load spikes caused a longer backlog of requests. By detecting the sudden increase of queuing delay, {\proj} temporarily switched to CFS. This helped quickly consume the backlog from the global queue so that normal load coming after the spikes can be served by using {\proj}'s default FILTER pool. A direct effect is a smooth queuing delay curve (Figure~\ref{fig:delay}) and considerable reduction of turnaround times for about $50\%$ of function requests (Figure~\ref{fig:switch_cdf}). 
More importantly, Figure~\ref{fig:switch_cdf} demonstrates that CFS or FILTER policy alone ({\proj} with overload detection disabled) is not sufficient to handle transient overload; {\proj}'s hybrid strategy effectively combines the best of both policies to achieve minimum turnaround time. 

\vspace{-4pt}
\section{OpenLambda Evaluation}
\label{sec:ol}
\vspace{-2pt}

\subsection{End-to-End Efficiency}
\label{subsec:ol_e2e}
\vspace{-2pt}

We used {\bench} to generate a more comprehensive FaaS workload, which includes three function applications: {\bf Fibonacci sequence} ({\fib}), {\bf markdown generation} ({\mg}), and {\bf sentiment analysis} ({\sa}).
\added{As mentioned, {\fib} calculates a sequence of $N$ Fibonacci numbers and is CPU-heavy. . 
{\mg} reads a JSON file from the function's local storage and transfers it to the markdown format; {\mg}'s execution is I/O-intensive. 
{\sa} reads a file that contains a sentiment vocabulary dictionary and then predicts the sentimentality given a target sentence; {\sa} is both CPU-intensive and I/O-intensive.}  
This workload reused the same function duration distribution and IAT distribution of the Azure-sampled workload. 
\added{OpenLambda was deployed to use 72 cores of the EC2 bare-metal instance, following an architecture depicted in Figure~\ref{fig:impl}. We varied the IAT to generate three load levels: $80\%$, $90\%$, and $100\%$. The OpenLambda deployment introduced extra overhead at various levels, including the OpenLambda worker servers and the HTTP sandbox servers. These overheads diminished the performance benefits of {\proj} to some extent; however, as we will show, {\proj} can still provide huge performance improvement for the majority of functions that are short.}

\begin{figure*}[h]
\begin{minipage}{\textwidth}
\begin{center}
\begin{minipage}[b]{0.37\textwidth}
\includegraphics[width=1\textwidth]{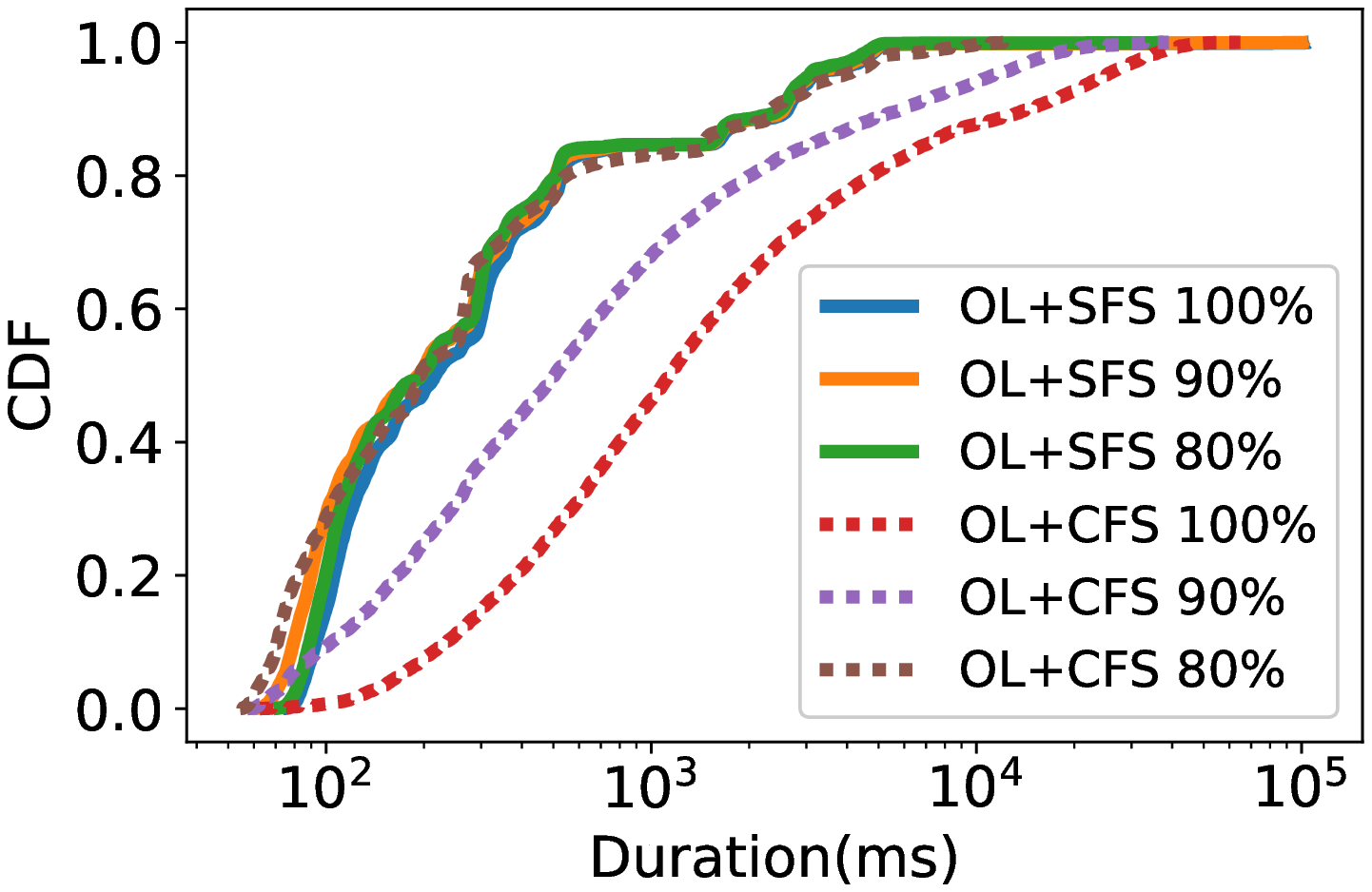}
\caption{Performance CDF.}
\label{fig:ol_cdf}
\vspace{10pt}
\end{minipage}
\hspace{30pt}
\begin{minipage}[b]{0.37\textwidth}
\includegraphics[width=1\textwidth]{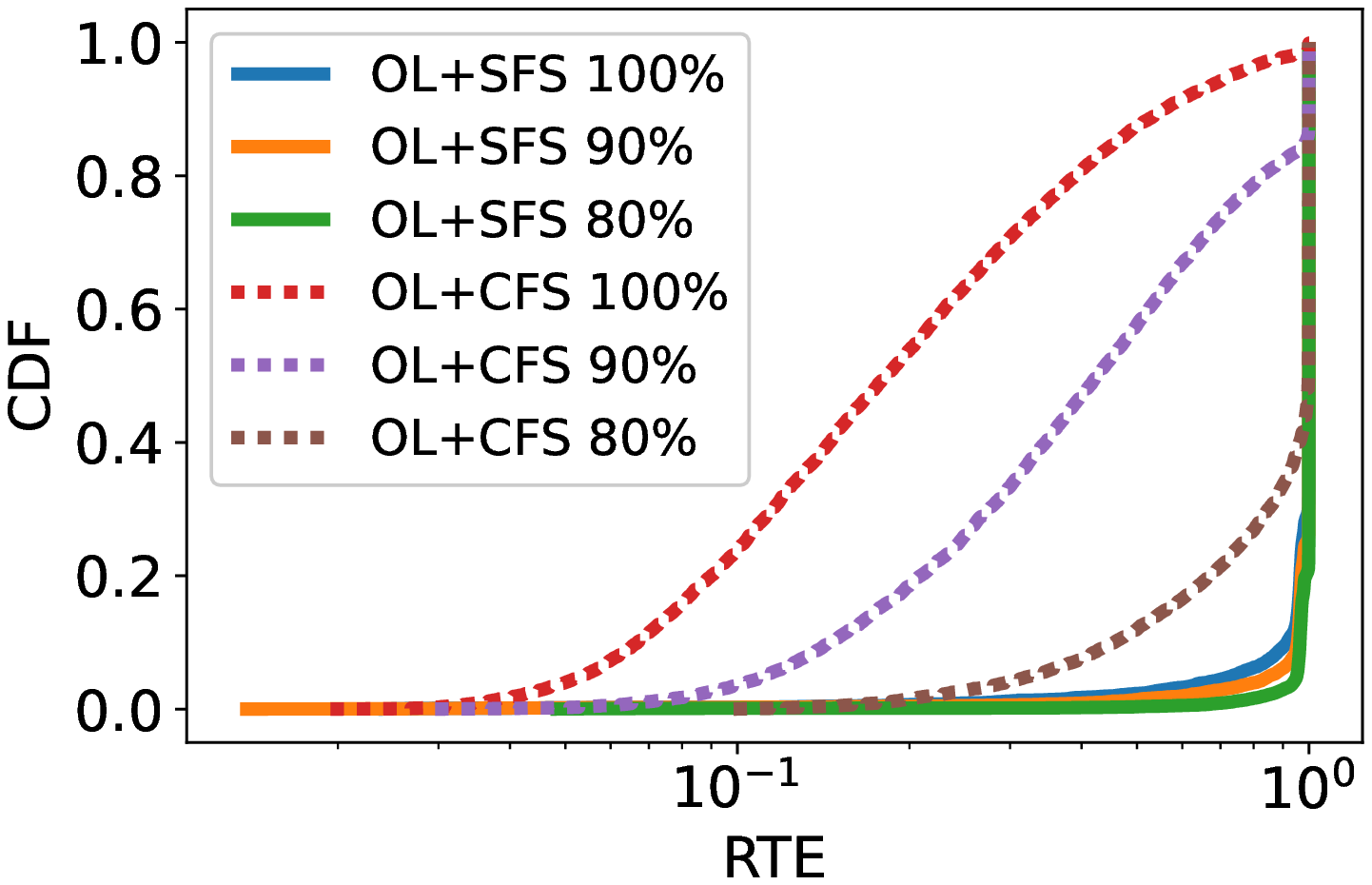}
\caption{RTE CDF.}
\label{fig:ol_rte}
\vspace{10pt}
\end{minipage}
\end{center}
\end{minipage}
\vspace{-15pt}
\end{figure*}

\begin{figure*}[h]
\begin{minipage}{\textwidth}
\begin{center}
\begin{minipage}[b]{0.37\textwidth}
\includegraphics[width=1\textwidth]{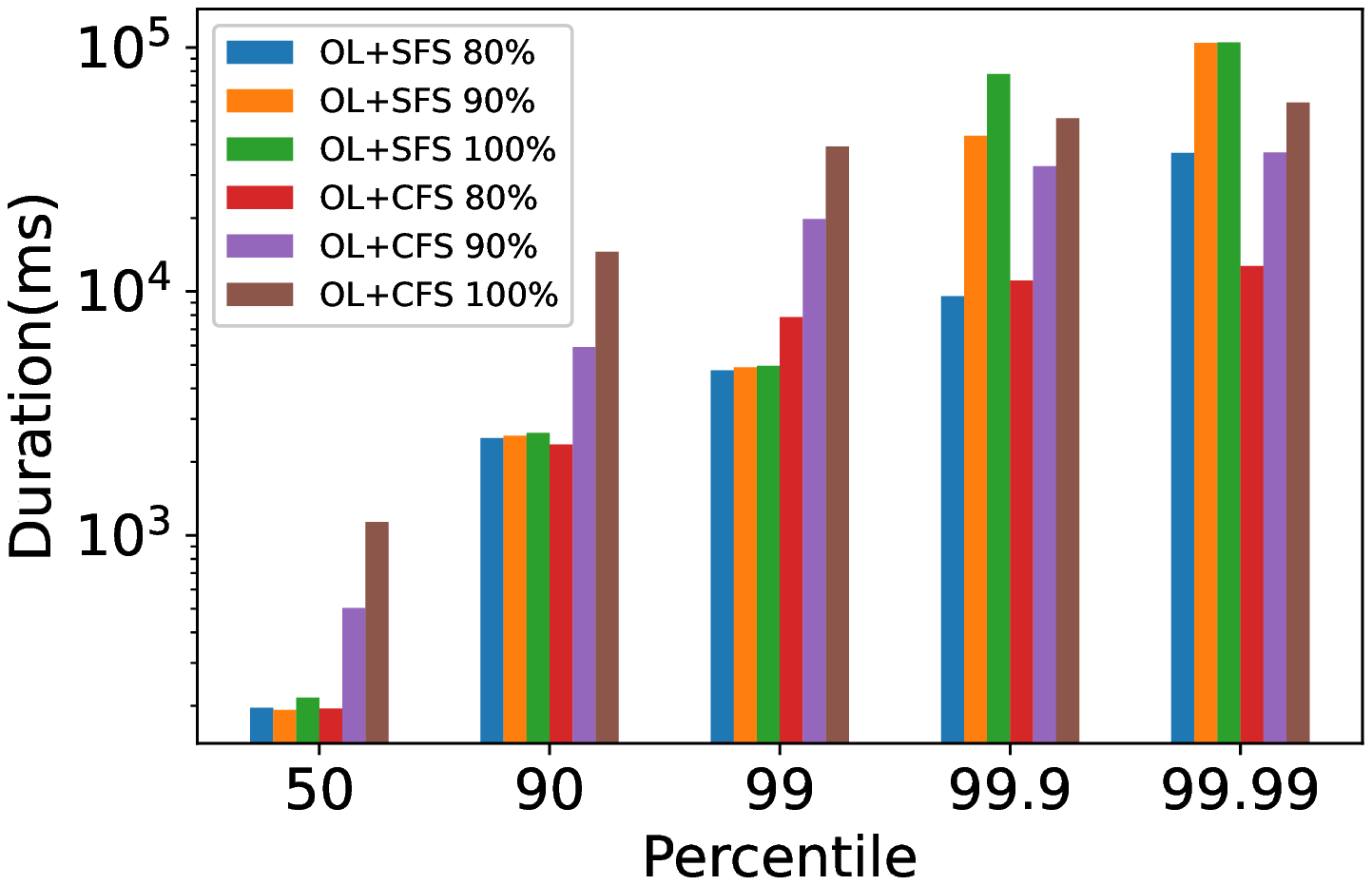}
\caption{
Percentile breakdowns of function execution duration.}
\label{fig:ol_percentile}
\vspace{-10pt}
\end{minipage}
\hspace{30pt}
\begin{minipage}[b]{0.37\textwidth}
\includegraphics[width=1\textwidth]{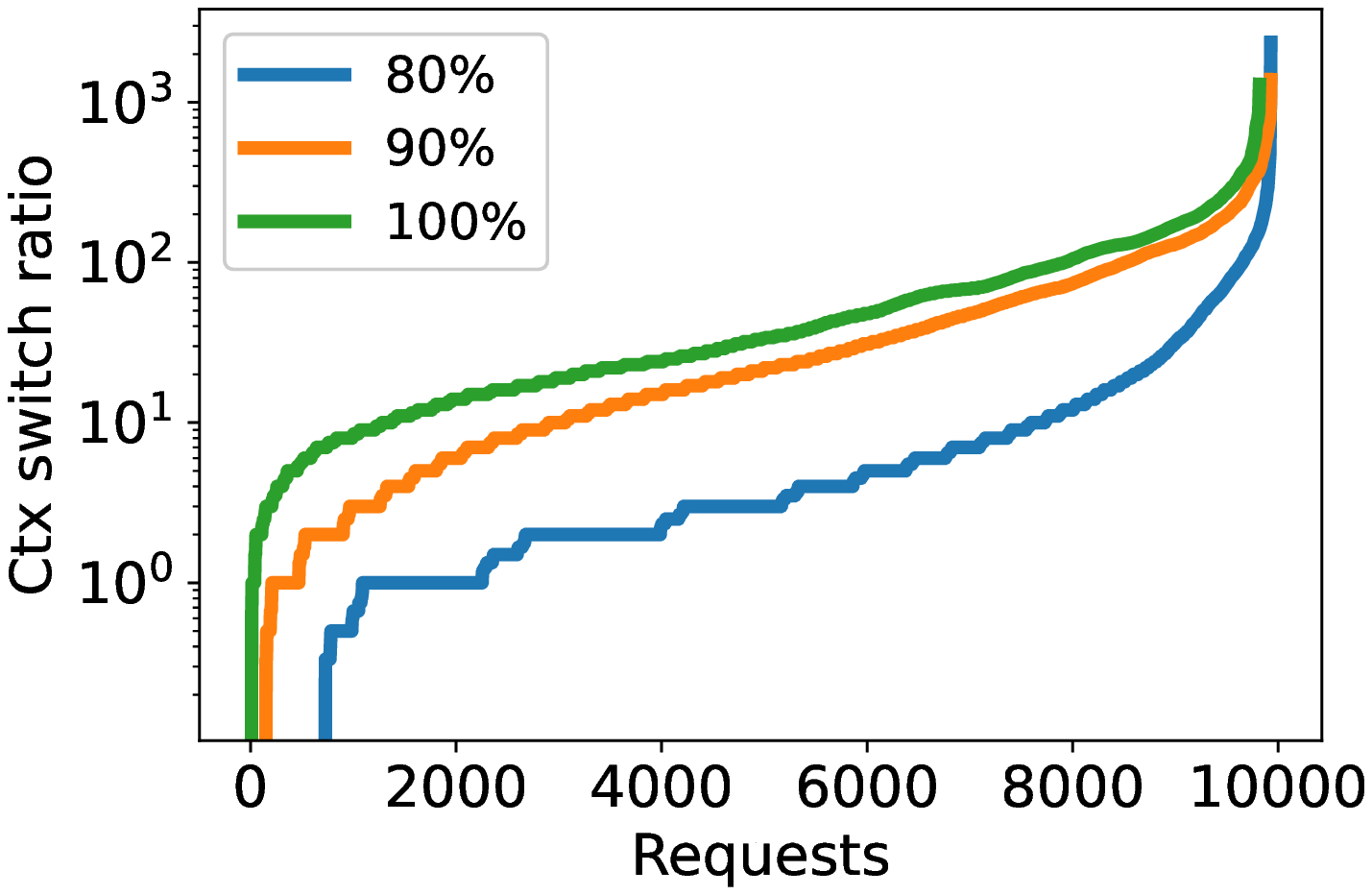}
\caption{The ratio of CFS context switches to that of {\proj}. 
}
\label{fig:ol_ctx}
\vspace{-10pt}
\end{minipage}
\end{center}
\end{minipage}
\end{figure*}

Figure~\ref{fig:ol_cdf} and \ref{fig:ol_rte} report the distributions of function execution duration and RTE. 
\added{The functions ran on average $14.1\%$ longer with OpenLambda+CFS under $80\%$ load than OpenLambda+{\proj} under the same load. When the load increased, OpenLambda+CFS started to see performance degradation, while OpenLambda+{\proj} achieved almost identical performance under all the three loads. As shown in Figure~\ref{fig:ol_percentile}, 
OpenLambda+{\proj} observed a $99^{th}$ percentile duration of $4.75$~seconds,
a $1.65\times$, $4.04\times$, and $7.93\times$ speedup compared to OpenLambda+CFS under the load of $80\%$, $90\%$, and $100\%$, respectively.}
\added{We also measured the number of context switches occurred under the three loads. Figure~\ref{fig:ol_ctx} shows the normalized context switches
for each function request. Under the $80\%$ and $100\%$ load, more than $99\%$ of function requests scheduled by CFS had more context switches than {\proj}. For about $85\%$ of requests, CFS suffered $10\times$ more context switches than {\proj}.}

\vspace{-4pt}
\subsection{{\proj} Overhead}
\label{subsec:ol_overhead}

{\proj} incurs a small runtime overhead. There are two sources of overhead: 
(1)~{\proj} uses goroutines as scheduling workers: function scheduling incurs some overhead;
(2)~{\proj} workers perform periodic polling to check the kernel status of the function process. The polling overhead is the dominant overhead.
\begin{wraptable}{r}{0.72\columnwidth}
\vspace{-5pt}
\caption{
{\proj}' (relative) CPU overhead in support of a 72-core OpenLambda deployment. 
}
\centering
\scalebox{0.8}{
\begin{tabular}{lrrrr}
\Xhline{2\arrayrulewidth}
{\bf Interval} & {\bf min} & {\bf average} & {\bf medium} & {\bf max}\\
\hline
{1~ms} & {1.6\%} & {3.8\%} &{3.8\%} &{6.2\%} \\
{4~ms} & {1.3\%} & {3.6\%} &{4.0\%} &{6.2\%}\\
{8~ms} & {1.4\%} & {3.4\%} &{3.9\%} &{6.6\%}\\
\Xhline{2\arrayrulewidth}
\end{tabular}
} 
\label{tbl:overhead}
\vspace{-5pt}
\end{wraptable}
Table~\ref{tbl:overhead} shows {\proj}' CPU usage in  the 72-core OpenLambda tests. With a polling \added{interval of 4~ms, {\proj}' average CPU usage was $259.8\%$ for the Azure-sampled workload, meaning that an extra of $2.6$ cores were needed in order to boost a 72-core OpenLambda deployment, a relative overhead of only $2.6 / 72 = 3.6\%$.} 
About $74.4\%$ of the total overhead was for periodic status polling, while the rest of $25.6\%$ was for scheduling activities.

%% file: discussion.tex
\section{Discussion}
\label{sec:discussion}

In this section, we discuss the limitations and possible future directions of {\proj}.

\noindent\textbf{Impact of Function Cold Start.}
Significant function cold start costs may offset the benefit of {\proj}, especially for short functions. Optimizing the cold start cost of serverless functions is an important and challenging problem that has drawn great attention from the community. Commercial FaaS platforms use sandbox and runtime caching extensively to mitigate the impact of cold start on function performance~\cite{infinicache_fast20, peeking_atc18}. The Azure Functions workload analysis~\cite{serverless_in_the_wild} reports that even a naive keep-alive function warmup policy can guarantee zero cold start for around $50\%$ of the function applications; with even smarter policies~\cite{serverless_in_the_wild, faascache_asplos20, icebreaker_asplos22},
the cold start rate could be further reduced to less than $10\%$ for all the function requests served on a single function host server. We foresee that most if not all the function requests would be executed without a cold start penalty with the recent advancement in cold start optimization~\cite{catalyzer_asplos20, container_sharing_atc22, fireworks_eurosys22, agile_cold_start_hotcloud19, sock_atc18, faasnet_atc21}; this makes the OS-level function scheduling---the ``last mile'' of a function request---a practical and urgent research problem that demands effective solutions like {\proj}.

\noindent\textbf{Why User-Space?}
{\proj} is designed to be a standalone, user-space function scheduler, which can be transparently plugged into existing FaaS platforms. While a kernel implementation of {\proj} would certainly work, with possibly less runtime overhead but much higher engineering efforts, a user-space implementation offers future-proof flexibility by retaining all the desirable properties of existing Linux scheduling facilities. With decades of research in datacenter workload co-location~\cite{borg_eurosys15, borg2_eurosys20, alibaba_colocation_socc18, google_trace_socc12, medea_eurosys18}, soon we will see co-location of production FaaS workloads with other cloud computing workloads. CFS, as the battle-tested, general-purpose scheduling solution for a wide range of workloads, 
would still play a key role in balancing the CPU resource usage. {\proj} is designed to co-exist with and complement an existing OS scheduler in these scenarios. 
Moreover, co-location of highly diverse workloads is likely to cause more intense CPU contention, thus demanding future research.

%% file: related.tex
\vspace{-4pt}
\section{Related Work}
\label{sec:related}
\vspace{-2pt}

\noindent\textbf{Scheduling Short and Long Jobs.}
Improving turnaround time by approximating SRTF is a well-known approach that has been investigated in many domains~\cite{overload_tit06, mlfq_sigmetrics95, bsd_os_txt, size_sched_tocs03, srtf_mor01}. 
A series of systems use request sizes as the hint to approximate SRTF. 
Size-based scheduling gives preference to requests for small files targeting web servers serving static HTTP requests~\cite{size_sched_tocs03}. 
Similarly, Harchol-Balter~et~al. applied SRTF to webserver request scheduling based on sizes of Linux kernel socket buffers~\cite{srtf_mor01}. Inspired by these works, {\proj}  presents a practical priority scheduler that addresses many of the challenges in emerging, real-world FaaS workloads.

\noindent\textbf{Scheduling for Fine-grained I/O Workloads.}
\added{Shenango~\cite{shenango_nsdi19}, Shinjuku~\cite{shinjuku_nsdi19}, and ZygOS~\cite{prekas2017zygos} use scheduling techniques such as core re-allocation, preemption, and work-stealing.
These techniques optimize tail latency of small key-value requests whose service time is highly predictable; Shenango and Shinjuku assume  long jobs co-located with small key-value request serving jobs---batch applications or range queries---whose application type is either known ahead or can be obtained from packet inspection. In contrast, {\proj} does not assume \emph{a priori} knowledge about function types or execution time but instead requires a very small amount of historical statistics for online time slice adjustment.}

\noindent\textbf{Serverless Function Scheduling.}
Centralized, core-granular scheduling~\cite{granular_socc19} uses two-level scheduling: it uses centralized scheduling to eliminate queue imbalance and core granular scheduling
to reduce the interference caused by proportional-share.
Core-granular scheduling assumes: 
(1)~non-preemption, meaning a function, once scheduled to a worker core, runs to completion (i.e., running in FIFO),
and (2)~massive distributed resources, meaning the scheduler can always find available cores to schedule a function request.
{\proj} shares similar goals but targets a local server scheduling environment, where the OS scheduling plays a critical role. 
Another line of work is focused on distributed or FaaS platform-level function scheduling~\cite{ensure_acsos20, fnsched_wosc19, nightcore_asplos21, olsched_minmove18, faasrank_acsos21} by using function placement optimization, low-latency I/Os, data locality, and reinforcement learning. 
Serverless dataflow frameworks use variants of cluster scheduling techniques~\cite{wukong_pdsw19, wukong_socc20, pywren_socc17, numpywren_socc20, sequoia_socc20} for serverless workflow applications. These works use the Linux scheduler for ``last mile'', OS-level task scheduling and would benefit from {\proj}.

\noindent\textbf{User-defined Scheduling.}
\added{Syrup~\cite{syrup_sosp21} and ghOSt~\cite{ghost_sosp21} allow developers to implement application-specific scheduling policies directly in the user space. Syrup uses the eBPF~\cite{ebpf} maps data structure to support user-kernel communication, while ghOST uses message queues and transactions for user-kernel communication. A user-defined policy may, however, observe significant user-kernel communication cost if the application needs frequent user-level scheduling adjustment; this is the case for serverless scheduling, where the time slice needs to be frequently and dynamically tuned by the scheduler.}

%% file: conclusion.tex
\vspace{-4pt}
\section{Conclusion}
\label{sec:conclusion}
\vspace{-2pt}

Serverless computing is gaining increasing popularity, as it promises 
fine-grained resource management, accounting, and billing at the milliseconds level.  
However, in practice,
FaaS workloads are highly heterogeneous and latency-sensitive, and have shown great volatility in execution durations. 
In this work, we have shown, via the design and implementation of a user-space scheduler {\proj} and empirical evaluation, that {\proj}, by approximating SRTF scheduling, can significantly reduce the execution duration of short functions. 
{\proj} approximates SRTF
with a dynamic and adaptive time slice in a first-level, global queue to combine the best worlds of FIFO and RR, while defaulting to the underlying OS-level scheduler in the second-level queue. 
{\proj} is transparent and can be easily ported to existing FaaS platforms. 
as we have demonstrated through an open-source FaaS platform OpenLambda. 
We hope that {\proj} will inspire new OS-level scheduling policies attuned to FaaS applications and open doors to new, FaaS-oriented SLO rules.

{\proj} and {\proj}-ported OpenLambda are available at:
\begin{center}
\vspace{-5pt}
\textbf{\url{https://github.com/ds2-lab/SFS}}.
\end{center}